\documentclass[11pt]{article}

\usepackage[final]{acl}

\usepackage{times}
\usepackage{latexsym}
\usepackage{float}
\usepackage[normalem]{ulem}
\usepackage{booktabs}
\usepackage{multirow}
\usepackage{threeparttable}
\usepackage[table]{xcolor}
\usepackage{amsmath}
\usepackage[linesnumbered,ruled,vlined]{algorithm2e}
\usepackage{array}
\usepackage{graphicx}
\usepackage{arydshln} 
\usepackage{dashrule}
\usepackage{makecell}
\usepackage{subcaption}
\usepackage{amssymb}
\usepackage{listings}
\usepackage{caption}
\usepackage{xcolor}
\usepackage{dsfont}
\usepackage{placeins}
\usepackage{CJKutf8}
\newcommand{\cn}[1]{\begin{CJK*}{UTF8}{gbsn}#1\end{CJK*}}
\usepackage[T1]{fontenc}
\usepackage[utf8]{inputenc}

\usepackage{microtype}

\usepackage{inconsolata}

\usepackage{graphicx}
\usepackage{arydshln}

\title{TED-TTS: Training-Free Intra-Utterance Emotion and Duration Control for Text-to-Speech Synthesis}

\author{
 \textbf{Qifan Liang\textsuperscript{1}\thanks{Equal contribution.},}
 \textbf{Yuansen Liu\textsuperscript{1}\footnotemark[1],}
 \textbf{Ruixin Wei\textsuperscript{1}\footnotemark[1],}
 \textbf{Nan Lu\textsuperscript{1},}
 \textbf{Junchuan Zhao\textsuperscript{1},}
 \textbf{Ye Wang\textsuperscript{1}}
\\
\textsuperscript{1}School of Computing, National University of Singapore
\\
\{liangqifan,yuansen,ruixin2003,nlu\}@u.nus.edu,
\{junchuan,wangye\}@comp.nus.edu.sg
}

\begin{document}
\maketitle

\begin{abstract}

While controllable Text-to-Speech (TTS) has achieved notable progress, most existing methods remain limited to inter-utterance-level control, making fine-grained intra-utterance expression challenging due to their reliance on non-public datasets or complex multi-stage training. In this paper, we propose TED-TTS, a training-free controllable framework for pretrained zero-shot TTS to enable intra-utterance emotion and duration expression. Specifically, we propose a segment-aware emotion conditioning strategy that combines causal masking with monotonic stream alignment filtering to isolate emotion conditioning and schedule mask transitions, enabling smooth intra-utterance emotion shifts while preserving global semantic coherence. Based on this, we further propose a segment-aware duration steering strategy to combine local duration embedding steering with global EOS logit modulation, allowing local duration adjustment while ensuring globally consistent termination. To eliminate the need for segment-level manual prompt engineering, we construct a 30,000-sample multi-emotion and duration-annotated text dataset to enable LLM-based automatic prompt construction. Extensive experiments demonstrate that our training-free method not only achieves state-of-the-art intra-utterance consistency in multi-emotion and duration control, but also maintains baseline-level speech quality of the underlying TTS model. Code\footnote{Code: \url{https://github.com/Simon-leong/TED-TTS}.} and audio samples\footnote{Audio samples: \url{https://simon-leong.github.io/TED-TTS-DemoPage/}.} are available.

\end{abstract}

\section{Introduction}

\begin{figure}[htbp]
    \centering
    \includegraphics[width=\columnwidth]{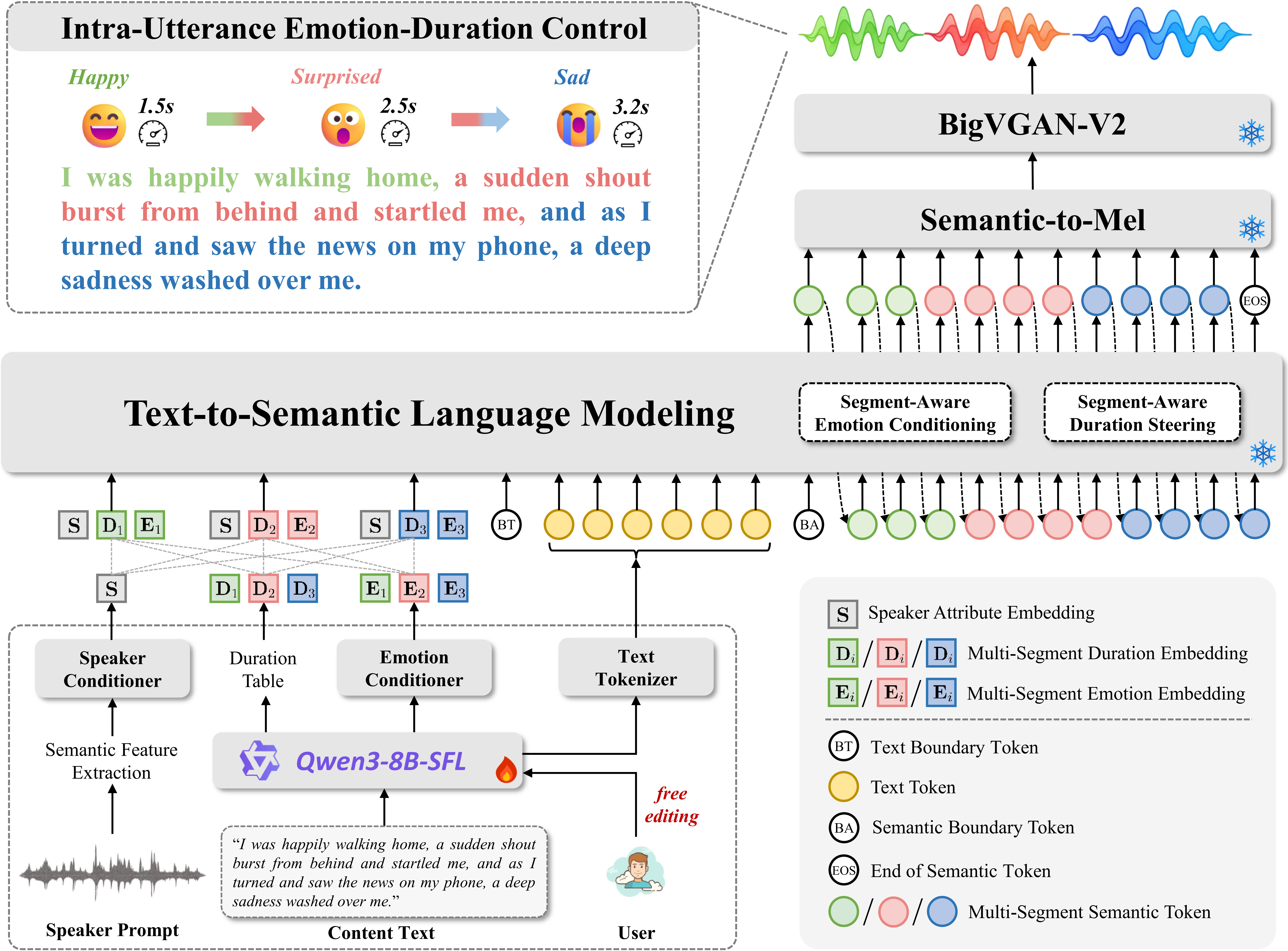}
    \caption{Overview of our training-free framework for intra-utterance emotion and duration control, where the green, red, and blue regions denote three segments with different emotion and duration settings within the same utterance.}
    \label{Intro}
\end{figure}

Humans naturally regulate emotional expression and speaking pace during speech in a dynamic and flexible manner, reflecting changes in semantics, emphasis, and discourse intent \cite{scherer2003vocal,tang2023squad}. How to replicate such intra-utterance expressiveness remains a challenge in building human-like TTS synthesis systems.

Recent advances in controllable TTS have enabled zero-shot synthesis conditioned on attributes such as speaker identity, emotion, and speaking rate \cite{du2024cosyvoice2,wang2025maskgct,wang2025spark,chen2025f5,gao2025emo,yang2025emovoice,zhou2025indextts2}. Despite these advances, controllability in most existing methods remains confined to the utterance level, where a single emotional or prosodic condition is uniformly applied to an entire utterance, deviating from the dynamic expression naturally observed in human speech. To address this limitation, some methods \cite{luo2024emotion,tan2024naturalspeech} predict phoneme- or frame-level affective attributes directly from text, while others \cite{kanda2024making,wu2024laugh} rely on emotional reference speech to guide localized expressive patterns, such as brief laughter or crying. Most recently, WeSCon \cite{wang2025word} proposes a self-training framework with transition smoothing and emotional-bias mechanisms, enabling the TTS model to render multiple emotions within an utterance through distillation. While these meaningful progress, they typically rely on large-scale time-aligned annotated speech datasets or involve multi-stage training pipelines, which substantially limit their cross-model transferability and real-world deployment.

These challenges naturally raise an important question: \textbf{\emph{Is it possible to achieve stable segment-level emotion transitions and duration control without retraining the model?}} In this paper, as shown in Fig.~\ref{Intro}, we revisit controllable TTS from an inference-time perspective and propose the first \textbf{T}raining-free Intra-Utterance \textbf{E}motion and \textbf{D}uration control framework (TED-TTS). Rather than introducing additional predictors or retraining the acoustic model, our approach focuses on restructuring how conditioning information is accessed and updated during autoregressive decoding. Specifically, for multi-emotion control, we propose a segment-aware emotion conditioning strategy that combines causal masking with monotonic stream alignment algorithm, which jointly isolates segment-specific emotion conditioning and performs online text-semantic alignment to schedule mask transitions, enabling smooth intra-utterance emotion shifts while preserving global semantic coherence. To enable multi-duration control, we further propose a segment-aware duration steering strategy to incorporate local duration embedding steering with global EOS logit modulation, allowing segment-level pacing adjustment while ensuring globally consistent sequence termination. Besides, we construct a \textbf{M}ulti-\textbf{E}motion and \textbf{D}uration-annotated text dataset (MED-TTS) with 30,000 samples and fine-tune Qwen3-8B to enable LLM-based automatic prompt construction, thereby eliminating the need for segment-level manual segmentation and prompt engineering. Extensive experiments demonstrate that our method achieves state-of-the-art performance in stable intra-utterance multi-emotion transitions and duration control, while preserving the strong zero-shot synthesis capability of the underlying TTS model without any additional training. Our contributions are summarized as follows:

\begin{itemize}
    \item We propose a training-free controllable framework for intra-utterance-level TTS, and eliminate manual prompt engineering by constructing a 30,000-sample multi-emotion and duration-annotated text dataset for LLM-based automatic prompt construction.

    \item We propose a segment-aware emotion conditioning strategy to jointly isolate segment-specific emotion conditioning and perform online text-semantic alignment, enabling stable multi-emotion transitions within a single utterance.

    \item We propose a segment-aware duration steering strategy that achieves local segment duration control while preserving globally consistent sequence termination.

    \item Extensive experiments demonstrate that our training-free method not only achieves state-of-the-art intra-utterance consistency for multi-emotion and duration control, but also maintains the baseline-level speech quality of the underlying TTS model.
\end{itemize}

\section{Related Work}
\subsection{Emotionally Controllable TTS}

Emotion-controllable TTS methods can be broadly categorized by the modality of emotion prompts. 
\textbf{Speech-prompt-based methods} condition synthesis on reference emotional utterances and can transfer fine-grained affective cues such as intensity and prosody \cite{eskimez2024e2, du2024cosyvoice1, wang2025maskgct, wang2025spark, chen2025f5}, but their reliance on reference speech limits practical flexibility. 
In contrast, \textbf{Text-prompt-based methods} offer more flexible control, where early approaches rely on discrete emotion labels \cite{guo2023emodiff, kang2023zet, diatlova2023emospeech, tang2024ed, gao2025emo}, while recent methods adopt natural language emotion descriptions for richer and more continuous conditioning \cite{guo2023prompttts, liu2023promptstyle, yang2024instructtts, du2024cosyvoice2, yang2025emovoice, zhou2025indextts2}. 
However, these methods typically operate at the utterance level, assigning a single global emotion to an entire utterance and thus failing to capture intra-utterance emotional dynamics. 
To address this limitation, several \textbf{Intra-utterance control methods} predict fine-grained affective attributes directly from text \cite{im2022emoq, luo2024emotion, tan2024naturalspeech}, or incorporate emotional reference speech to enable localized expressions such as laughter or crying \cite{kanda2024making, wu2024laugh}. 
More recently, WeSCon \cite{wang2025word} introduces a self-training framework to support multi-emotion rendering within a single utterance. 
Despite these advances, existing methods often rely on large-scale non-public emotional datasets or multi-stage training pipelines that hinder their scalability and cross-model transferability, leaving training-free intra-utterance emotion control as an open and practically valuable challenge.

\subsection{Duration Controllable TTS}

Current exploration on duration control has advanced along both non-autoregressive and autoregressive approaches. \textbf{Non-autoregressive methods} achieve duration control via explicit duration predictors based on diffusion-transformers \cite{lee2025dittotts}, flows \cite{kim2023sc}, or language models \cite{du2025cosyvoice3}, but these predictors are trained separately and often struggle with temporal accuracy under prosodic variability. In contrast, \textbf{autoregressive methods} lack inherent duration control and typically rely on auxiliary cues, such as natural-language timing prompts \cite{zhou2024voxinstruct} or specialized attributes and labels \cite{li2025flespeech, sahipjohn2024dubwise, wang2025spark}. More recently, IndexTTS2 \cite{zhou2025indextts2} improves controllability by conditioning semantic token generation on duration positional embeddings, enabling more stable alignment between desired and produced token lengths than earlier autoregressive methods. However, existing approaches still struggle to decouple local pacing from global generation, failing to provide a unified framework that ensures stable intra-utterance duration control without compromising overall alignment.

\subsection{Inference-Time Controllable TTS}

Several approaches have explored inference-time controllable TTS, enabling flexible manipulation of speech attributes. 
EmoKnob \cite{chen-etal-2024-emoknob} injects scaled emotion difference vectors into speaker embeddings for emotion control, while PRESENT \cite{lam2025present} performs rule-based prosody shaping by adjusting pitch, duration, and energy predictions from text prompts. SPTTS \cite{suni2025style} further operates in the latent embedding space, manipulating prosody and style directions derived via linear regression and vector arithmetic. More recently, EmoSteer-TTS \cite{xie2025emosteer} directly steers token-level activations in pretrained diffusion-based TTS models, enabling training-free emotion control with improved interpretability over global embedding methods. However, these methods predominantly focus on implicit latent manipulation or isolated feature editing, and lack a unified framework for jointly controlling segment-level emotion and pacing transitions.

\section{Method}
In this section, we introduce a training-free controllable framework for intra-utterance emotion and duration transitions, with details provided in the following subsections.

\subsection{MED-TTS Dataset}
% Not finished yet.
\textbf{Automatic Prompt Construction.} Existing intra-utterance controllable TTS systems require manual text segmentation and segment-level emotion and duration specification, which is labor-intensive in real-world scenarios. To eliminate manual prompt engineering, we fine-tune the Qwen3-8B LLM to automatically transform raw user text into structured multi-segment prompts. As a prerequisite, we construct a dedicated \textbf{M}ulti-\textbf{E}motion and \textbf{D}uration-annotated text dataset (MED-TTS) with 30,000 samples, which is used to supervise emotion-aware text segmentation, natural language emotion description generation, and segment-level speech duration estimation. As illustrated below, MED-TTS is synthesized using LLM through a structured pipeline consisting of generation, annotation, and verification. 

\begin{figure*}[htbp]
    \centering
    \includegraphics[width=\textwidth]{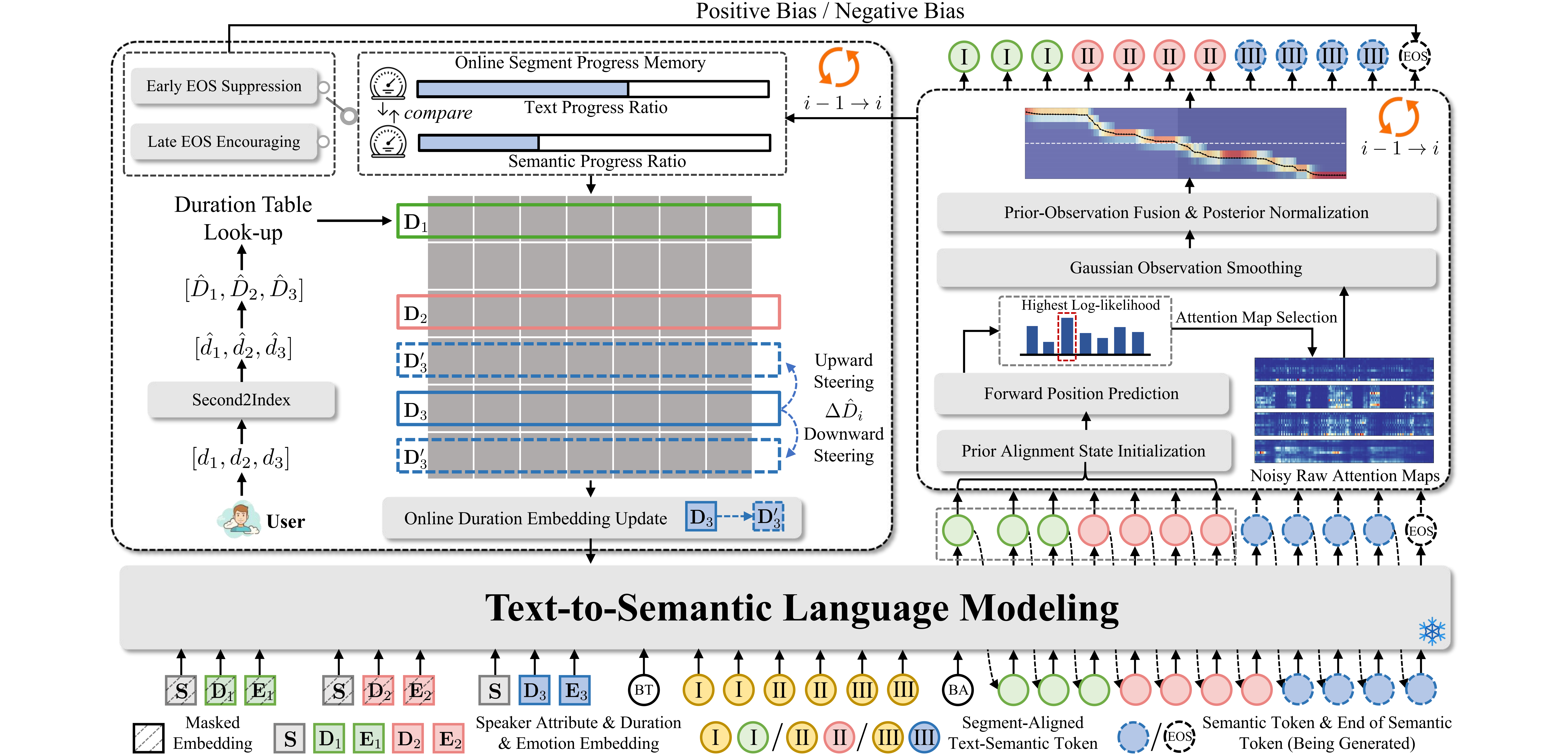}
    \caption{Overview of our training-free framework for fine-grained intra-utterance emotion and duration control, illustrating the transition from the second (red) segment to the third (blue) segment via segment-aware duration steering (left) and segment-aware emotion conditioning (right) strategy.}
    \label{method}
\end{figure*}

\textbf{Step 1: Content text generation.} 
GPT-4o is prompted to generate 15,000 English and 15,000 Chinese emotion-rich content texts that explicitly exhibit continuous emotional transitions
within a single utterance. Each text spans multiple emotional phases drawn from 7 core emotions (happy, sad, angry, surprised, fearful, disgusted, and neutral) and 3 text categories (descriptive, dialogue-style, and observational), forming either smooth or abrupt emotional progressions across diverse content genres. Example prompt used in Step~1 is shown in List.~\ref{lst1}.
\textbf{Step 2: Multi-segment prompt annotation.} 
To enable precise segment-level control, DeepSeek-Chat is prompted to perform precise semantic decomposition by segmenting the utterance
into multiple emotion-specific segments. For each segment, it produces a concise natural language emotion description together with a realistic duration estimate, forming a structured sequence
of emotion-duration pairs compatible with controllable TTS inputs. Example prompt used in Step~2 is shown in List.~\ref{lst2}.
\textbf{Step 3: Post-processing and manual verification.} Automatic checks are finally applied to filter samples with formatting errors, missing fields, or invalid segment boundaries, followed by systematic manual verification of outputs from both Step~1 and Step~2. The manual verification checklist is provided in List~\ref{lst3}.

\begin{figure}[t!]
    \centering
    \begin{subfigure}[b]{0.48\columnwidth}
      \begin{minipage}[c][4cm][c]{\linewidth}
        \centering
        \includegraphics[width=\textwidth, trim=40mm 3mm 40mm 3mm, clip]{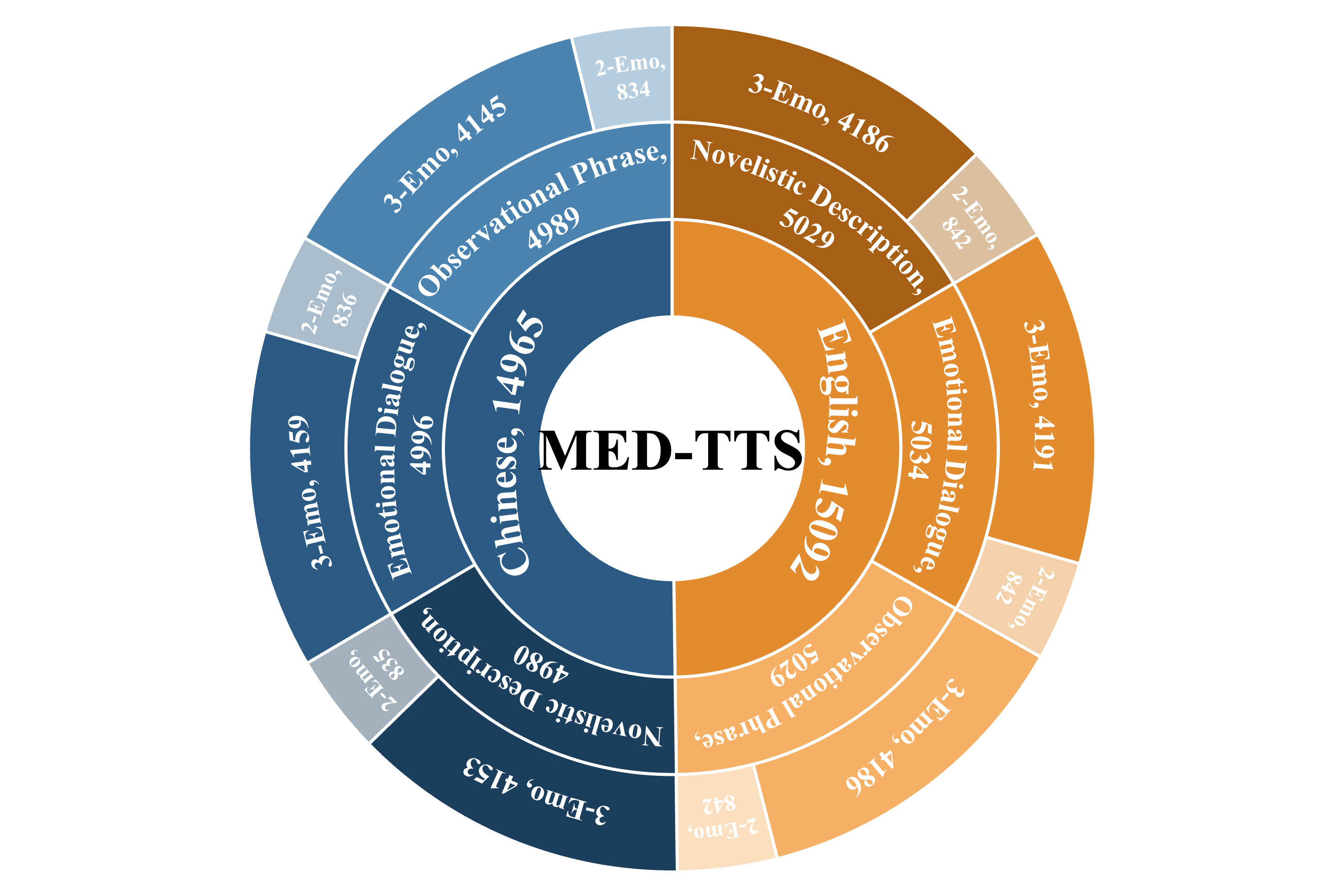}
      \end{minipage}
      \caption{Distribution of Content Segments}
      \label{dataset_method:a}
    \end{subfigure}
    \hfill
    \begin{subfigure}[b]{0.48\columnwidth}
      \begin{minipage}[c][4cm][c]{\linewidth}
        \centering
        \includegraphics[width=\textwidth]{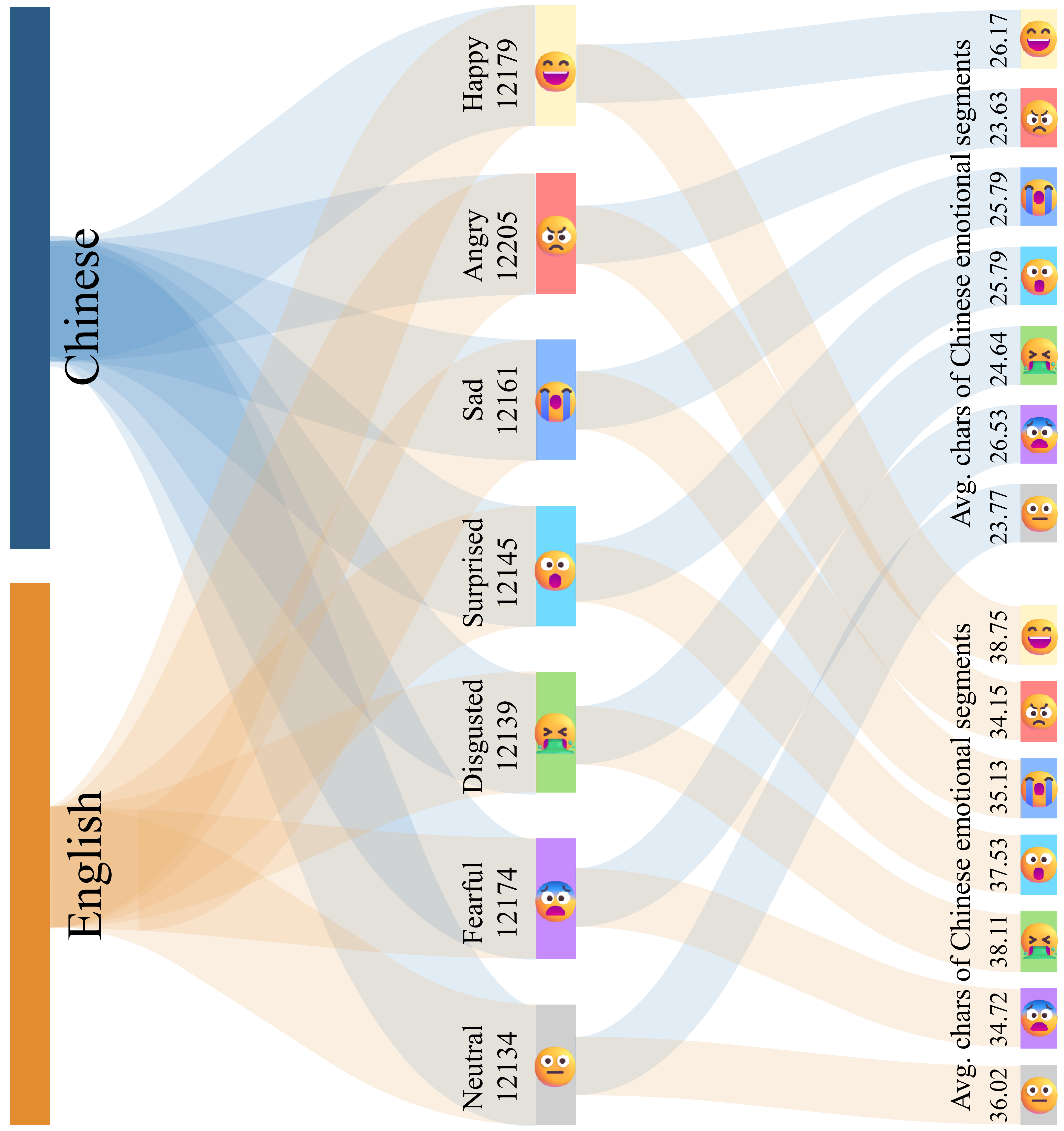}
      \end{minipage}
      \caption{Distribution of Emotion Segments}
      \label{dataset_method:b}
    \end{subfigure}
    \caption{Statistics of the MED-TTS dataset. (a) Distribution of Chinese and English content segments. (b) Distribution of emotion segments with average character counts.}
    \label{dataset_method}
\end{figure}

\textbf{Dataset Statistics}
We summarize the distribution of the MED-TTS dataset across languages, text categories, and emotion segments. As illustrated in Fig.~\ref{dataset_method:a}, the dataset is well balanced across languages, comprising 14,965 Chinese and 15,092 English samples. Within each language, samples are further evenly distributed across three text categories, each contributing approximately 5,000 utterances. Fig.~\ref{dataset_method:b} presents the segment-level emotion statistics. Across both Chinese and English, the 7 emotion types are uniformly represented, with each emotion accounting for approximately 12,000 segments. The average segment length remains stable within each language but differs across languages, with Chinese emotional segments typically spanning about 24-26 characters, while English segments are longer on average, ranging from roughly 34-39 words depending on emotion. Based on this dataset, we perform supervised fine-tuning with LoRA on the Qwen3-8B large language model, enabling automatic construction of segment-level TTS prompts without manual prompt engineering. Detailed prompting strategies, step-wise checklists, dataset statistics, and fine-tuning details are provided in the Appendix~\ref{appendix:A}.

\subsection{Segment-Aware Emotion Conditioning}
\label{subsec:multi_segment_control}

Our TTS architecture follows the same configuration as the IndexTTS2 \cite{zhou2025indextts2} baseline, and we focus our design on its text-to-semantic (T2S) module to enable training-free intra-utterance emotion control. Specifically, T2S is formulated as an autoregressive semantic token prediction task conditioned on text and a set of control embeddings.
Given an input text, represented by the yellow tokens in Fig.~\ref{method}, we decompose it into $M$ user-defined segments $\mathbf{X} = \{\mathrm{X}_1, \mathrm{X}_2, \dots, \mathrm{X}_M\}$, and each segment $\mathrm{X}_m$ is assigned a condition embedding $\mathbf{C}_m=\{\mathbf{I}, \mathbf{E}_m\}$, where $\mathbf{I}$ denotes a fixed speaker identity embedding shared across the segment, and $\mathbf{E}_m$ represents segment-specific emotion conditions. However, in autoregressive T2S formulations, semantic tokens are generated as a continuous stream without explicit segment boundaries, making it non-trivial to apply segment-level conditions to their corresponding text segments while preserving semantic continuity. To address this challenge, we propose a 2D causal attention mask combined with a monotonic stream alignment algorithm to enable smooth intra-utterance emotion transitions.

\textbf{2D Causal Attention Mask.}
To resolve the misalignment between continuous generation and segment-level conditions, we design a 2D causal attention mask that disentangles condition visibility from semantic context. The mask preserves standard causal attention among text and semantic tokens across segment boundaries, ensuring globally coherent semantic generation while strictly restricting access to condition embeddings to be segment-local. Specifically, for any token that belongs to the $m$-th segment (either a text token in $\mathrm{X}_m$ or a generated semantic token that currently aligns to $\mathrm{X}_m$), attention is allowed to attend only to its corresponding condition embedding $\mathbf{C}_m$, while all other condition embeddings $\{\mathbf{C}_j \mid j \neq m\}$ are masked out. Meanwhile, each condition embedding $\mathbf{C}_m$ is prevented from attending to other condition embeddings, avoiding cross-condition information leakage. After that, as shown at the bottom of Fig.~\ref{method}, emotional style is governed exclusively by the locally active condition, whereas semantic content remains globally visible through standard causal context. 

However, applying 2D causal attention masks requires real-time knowledge of the alignment between generated semantic tokens and source text tokens. While transformer attention can provide alignment cues, raw attention maps are often noisy, head-dependent, and non-monotonic, making them unreliable for driving mask transitions. To address this, we propose an online \textbf{M}onotonic \textbf{S}tream \textbf{A}lignment (MSA) algorithm that performs Bayesian-style alignment tracking using attention as observation.

\begin{figure}[t!]
    \centering
    \includegraphics[width=\columnwidth]{images/MSA.png}
    \caption{Detailed illustration of Monotonic Stream Alignment (MSA) in segment-aware emotion conditioning, where from top to bottom are MSA algorithm, MSA alignment result, and the visualization of 2D causal attention mask, respectively.}
    \label{msa}
\end{figure}

\textbf{Monotonic Stream Alignment (MSA).}
As shown in Fig.~\ref{msa}, we use $\mathbf{A}_i\in\mathbb{R}^{L\times H\times T}$ denote the raw attention maps from the current semantic token $\mathbf{s}_i$ to the $T$ text tokens across $L$ layers and $H$ heads, where $\mathbf{A}_i^{(l,h)}$ is the attention vector of head $(l,h)$. During online autoregressive decoding, MSA maintains a belief distribution over text positions to track the alignment of $\mathbf{s}_i$, represented by a prior distribution $\hat{\boldsymbol{\pi}}_i$ and a posterior distribution $\boldsymbol{\pi}_i$, both defined over the $T$ text tokens. At each decoding step $i$, MSA first performs the \emph{Predict} step by propagating the posterior $\boldsymbol{\pi}_{i-1}$ from the previous step forward along the text sequence using a monotonic transition operator $\mathcal{P}$. This propagation yields a prior distribution $\hat{\boldsymbol{\pi}}_i$ that encodes strong temporal monotonicity, encouraging gradual forward movement while suppressing backward alignment. After obtaining the monotonic prior $\hat{\boldsymbol{\pi}}_i$, MSA enters the \emph{Select} step to select the most reliable attention head by measuring how well each head’s attention distribution agrees with the predicted alignment:  

\begin{equation}
(l^\ast, h^\ast)
= \operatorname*{arg\,max}_{l,h}
\hat{\boldsymbol{\pi}}_i^\top \log \mathbf{A}_i^{(l,h)}, 
\end{equation}

\noindent where $\mathbf{A}_i^{(l,h)}$ denotes the attention vector of head $(l,h)$. The resulting head $(l^\ast,h^\ast)$ provides the most reliable attention observation used in the subsequent update. In the final \emph{Update} step, MSA combines the selected attention observation $\mathbf{A}_i^{(l^\ast,h^\ast)}$ with the monotonic prior $\hat{\boldsymbol{\pi}}_i$ to compute the posterior alignment belief as:

\begin{equation}
\boldsymbol{\pi}_i
= \frac{\hat{\boldsymbol{\pi}}_i \odot \mathcal{G}_\sigma\!\left(\mathbf{A}_i^{(l^\ast,h^\ast)}\right)}{Z}, 
\end{equation}

\noindent where $\odot$ denotes element-wise multiplication, $\mathcal{G}_\sigma(\cdot)$ is a Gaussian smoothing operator, and $Z$ is a normalization factor. This update incorporates real-time attention evidence while enforcing monotonicity, resulting in a stable alignment trajectory $\boldsymbol{\pi}_i$. Benefiting from this alignment trajectory, segment-level causal mask switching is triggered by tracking the expected aligned text position, enabling subsequent semantic tokens to attend to the new segment condition. More detailed mathematical derivations are provided in Appendix~\ref{appendix:B}.

\subsection{Segment-Aware Duration Steering}
\label{subsec:multi_segment_duration_control}
Beyond segment-level emotional expressiveness, we further extend our emotion control framework to enable multi-segment duration control in a fully training-free autoregressive setting.

\noindent\textbf{Local Duration Embedding Steering.}
Inspired by IndexTTS2~\cite{zhou2025indextts2}, we condition duration control on a dedicated duration embedding indexed by the semantic token length, and tie its embedding table $\mathbf{W}_{dur}$ with the semantic positional embedding table $\mathbf{W}_{sem}$ to align autoregressive positional progression with target duration. As shown in Fig.\ref{method}, given an utterance with $M$ segments and desired durations $\mathbf{d} = \{d_1, d_2, \dots, d_M\}$, each segment duration is converted into the corresponding number of semantic tokens according to the codec token rate~\cite{wang2025maskgct}, yielding $\hat{\mathbf{d}} = \{\hat{d}_1, \hat{d}_2, \dots, \hat{d}_M\}$.  
We accumulate segment-level targets into cumulative token lengths $\hat{D}_i=\sum_{k=1}^{i}\hat{d}_k$ and retrieve segment-wise initial duration embeddings as $\mathbf{D}_i=\mathbf{W}_{dur}[\hat{D}_i]$, which are concatenated into the segment-level conditioning inputs $\mathbf{C}_m$ to guide subsequent generation.

During autoregressive decoding, the actual semantic token generation speed may deviate from the user-specified target due to alignment uncertainty and model stochasticity. To correct such deviations online, we introduce a local duration embedding steering mechanism that dynamically updates the duration embedding via adaptive duration table lookup.  
At each decoding step $i$, we leverage MSA (Section~\ref{subsec:multi_segment_control}) to estimate the current aligned text position and compute two normalized progress indicators within the active segment: text progress $r_{text}$ and semantic progress $r_{sem}$.  
Their discrepancy is defined as $\Delta r = r_{text} - r_{sem}$, where a positive value indicates lagging semantic generation, which is then used to adjust the effective semantic token length via a proportional controller:

\begin{equation}
\Delta \hat{D}_{i} = \mathrm{clip}\!\left(\left\lfloor k \cdot \Delta r \right\rceil,\; -\Delta_{\max},\; \Delta_{\max}\right),
\end{equation}

\noindent where $k$ controls the correction strength, $\lfloor\cdot\rceil$ denotes rounding to the nearest integer, and $\Delta_{\max}$ bounds the maximum adjustment. The effective segment-wise target is updated as $\hat{D}_{i} + \Delta \hat{D}_{i} \rightarrow \hat{D}^{\prime}_{i}$, and the duration table $\mathbf{W}_{dur}$ is re-queried only for the active segment to obtain the updated duration embedding $\mathbf{D}^{\prime}_{i}$, while duration embeddings of other segments remain unchanged.  
For stability, updates are applied at a low temporal frequency, allowing multiple consecutive semantic tokens to share the same duration embedding.

\noindent\textbf{Global EOS Steering.}
In autoregressive decoding, the End-Of-Semantic (EOS) token determines sequence termination and overall duration. While local duration embedding steering regulates local generation pace, it does not explicitly control when decoding ends. To address this, we introduce a global EOS steering strategy that modulates sequence termination by applying adaptive biases to the EOS logit. Specifically, EOS generation is suppressed in all non-final segments to prevent premature termination, and in the final segment, the EOS logit is progressively adjusted based on the remaining semantic budget, discouraging early termination while smoothly encouraging EOS emission as the target budget is approached. Detailed parameter settings are provided in Appendix~\ref{appendix:C}.

% ---------------------- Table 1: Speech & Text Emotion Prompt ----------------------
\begin{table*}[t]
\centering
\resizebox{\textwidth}{!}{%
\begin{threeparttable}
\begin{tabular}{c l c c c c c c c c c c}
\toprule
 & \textbf{Model} & \textbf{WER/CER$\downarrow$} & \textbf{DNSM$\uparrow$} &
 \textbf{SSIM$\uparrow$} & \textbf{NISQA$\uparrow$} & \textbf{OVRL$\uparrow$} &
 \textbf{Emo2v$\uparrow$} & \textbf{SMOS$\uparrow$} &
 \textbf{NMOS$\uparrow$} & \textbf{EMOS$\uparrow$} \\
\midrule

\multicolumn{12}{l}{\textbf{\textit{Speech Emotion Prompt}}} \\
\midrule

% --- English Section (Speech) ---
 & MaskGCT & 3.520 & 3.829 & 0.347 & 4.475 & 3.275 & \underline{0.854}
 & 2.96$_{\pm0.34}$ & 2.77$_{\pm0.28}$ & \underline{3.64}$_{\pm0.24}$ \\
 & F5TTS & 2.632 & 3.674 & 0.353 & 4.427 & 3.330 & 0.832
 & 3.33$_{\pm0.36}$ & \underline{3.40}$_{\pm0.32}$ & 3.56$_{\pm0.28}$ \\
 & SparkTTS & \underline{2.433} & 3.456 & 0.358 & 4.494 & \textbf{3.404} & 0.849
 & \underline{3.49}$_{\pm0.29}$ & 3.27$_{\pm0.31}$ & 3.44$_{\pm0.29}$ \\
 & CosyVoice2 & \textbf{1.411} & 3.605 & 0.402 & \underline{4.535} & 3.316 & 0.831
 & 3.33$_{\pm0.31}$ & 2.87$_{\pm0.32}$ & 3.31$_{\pm0.28}$ \\
 & IndexTTS2 & 2.454 & \underline{3.871} & \underline{0.457} & 4.465 & 3.304 & \textbf{0.861}
 & 3.20$_{\pm0.36}$ & 2.98$_{\pm0.30}$ & \textbf{4.07}$_{\pm0.26}$ \\
\rowcolor{gray!15} \cellcolor{white}
\multirow{-6}{2.4em}{\centering\rotatebox[origin=c]{90}{English}}
 & \textbf{Ours} & 2.519 & \textbf{3.925} & \textbf{0.485} & \textbf{4.706}
 & \underline{3.395} & 0.837
 & \textbf{4.00}$_{\pm0.24}$ & \textbf{4.20}$_{\pm0.23}$ & 3.42$_{\pm0.30}$ \\

\addlinespace[0.4ex]
\cdashline{1-12}
\addlinespace[0.4ex]

% --- Chinese Section (Speech) ---
 & MaskGCT & 7.221 & 3.693 & 0.350 & 4.309 & 3.278 & \underline{0.814}
 & 2.80$_{\pm0.34}$ & 2.33$_{\pm0.31}$ & 3.64$_{\pm0.27}$ \\
 & F5TTS & 10.317 & 3.314 & 0.324 & 3.718 & 3.228 & 0.734
 & 3.22$_{\pm0.36}$ & 2.49$_{\pm0.38}$ & 3.13$_{\pm0.28}$ \\
 & SparkTTS & \textbf{3.107} & 3.466 & 0.382 & \underline{4.338} & \underline{3.345} & 0.807
 & 3.42$_{\pm0.33}$ & 2.87$_{\pm0.31}$ & \underline{3.80}$_{\pm0.24}$ \\
 & CosyVoice2 & \underline{3.375} & 3.306 & \underline{0.423} & 4.147 & 3.313 & 0.766
 & 3.04$_{\pm0.35}$ & 2.71$_{\pm0.37}$ & 3.29$_{\pm0.29}$ \\
 & IndexTTS2 & 4.015 & \underline{3.694} & 0.401 & 4.146 & 3.289 & \textbf{0.869}
 & \underline{3.67}$_{\pm0.33}$ & \underline{3.02}$_{\pm0.30}$ & \textbf{3.87}$_{\pm0.24}$ \\
\rowcolor{gray!15} \cellcolor{white}
\multirow{-6}{2.4em}{\centering\rotatebox[origin=c]{90}{Chinese}}
 & \textbf{Ours} & 3.792 & \textbf{3.752} & \textbf{0.470} & \textbf{4.509}
 & \textbf{3.370} & 0.724
 & \textbf{4.13}$_{\pm0.23}$ & \textbf{4.07}$_{\pm0.30}$ & 3.62$_{\pm0.32}$ \\

\midrule
\multicolumn{12}{l}{\textbf{\textit{Text Emotion Prompt}}} \\
\midrule

% --- English Section (Text) ---
 & CosyVoice2 & \textbf{1.522} & 3.465 & \underline{0.453} & \underline{4.330}
 & \underline{3.271} & 0.303
 & 3.33$_{\pm0.37}$ & \underline{3.73}$_{\pm0.31}$ & 2.53$_{\pm0.31}$ \\
 & IndexTTS2 & \underline{2.246} & \underline{3.543} & 0.424 & 4.299
 & 3.216 & \textbf{0.525}
 & \underline{3.76}$_{\pm0.39}$ & 3.44$_{\pm0.35}$ & \underline{3.42}$_{\pm0.29}$ \\
\rowcolor{gray!15} \cellcolor{white}
\multirow{-3}{1.8em}{\centering\rotatebox[origin=c]{90}{English}}
 & \textbf{Ours} & 3.038 & \textbf{3.694} & \textbf{0.462} & \textbf{4.569}
 & \textbf{3.335} & \underline{0.433}
 & \textbf{4.04}$_{\pm0.29}$ & \textbf{4.22}$_{\pm0.23}$ & \textbf{3.64}$_{\pm0.31}$ \\

\addlinespace[0.4ex]
\cdashline{1-12}
\addlinespace[0.4ex]

% --- Chinese Section (Text) ---
 & CosyVoice2 & \textbf{4.488} & 3.105 & \textbf{0.477} & \underline{4.346}
 & \underline{3.206} & 0.222
 & 2.18$_{\pm0.33}$ & \underline{3.56}$_{\pm0.31}$ & 2.84$_{\pm0.35}$ \\
 & IndexTTS2 & 6.962 & \underline{3.212} & 0.369 & 4.179
 & 3.169 & \textbf{0.702}
 & \underline{3.29}$_{\pm0.36}$ & 2.71$_{\pm0.33}$ & \underline{3.56}$_{\pm0.31}$ \\
\rowcolor{gray!15} \cellcolor{white}
\multirow{-3}{1.8em}{\centering\rotatebox[origin=c]{90}{Chinese}}
 & \textbf{Ours} & \underline{5.893} & \textbf{3.357} & \underline{0.421} & \textbf{4.407}
 & \textbf{3.295} & \underline{0.531}
 & \textbf{4.07}$_{\pm0.25}$ & \textbf{4.04}$_{\pm0.22}$ & \textbf{3.84}$_{\pm0.25}$ \\
\bottomrule
\end{tabular}
\end{threeparttable}
}
\caption{Objective and subjective evaluation across different emotion prompt settings. $\downarrow$ indicates that lower values are better, while $\uparrow$ indicates that higher values are better. Subjective results are evaluated by 15 listeners, with 95\% confidence intervals computed using a t-test. The best results are highlighted in \textbf{bold}, and the second-best results are \underline{underlined}.}
\label{com_1}
\end{table*}

\section{Experiments}

\noindent\textbf{Datasets and Comparison Models.}
We use the MED-TTS dataset for content text, text-based emotion prompts, and duration annotations, which contains 15,000 English and 15,000 Chinese pair samples. For each language, 500 samples are randomly held out for evaluation, while the remaining samples are used for Qwen3-8B fine-tuning. For identity and emotion speech prompts, we adopt the Emotional Speech Dataset (ESD) \cite{zhou2022emotional}, where a same-language speaker is fixed per test utterance to ensure timbre consistency, and the speaker’s emotional speech is used as segment-level emotion references.

We compare our method with representative controllable TTS methods spanning both non-autoregressive and autoregressive frameworks. The non-autoregressive models include MaskGCT \cite{wang2025maskgct} and F5-TTS \cite{chen2025f5}. The autoregressive models include CosyVoice2 \cite{du2024cosyvoice2}, Spark-TTS \cite{wang2025spark} and IndexTTS2 \cite{zhou2025indextts2}.

\noindent\textbf{Evaluation Metrics.}
We adopt both objective and subjective metrics to comprehensively evaluate system performance. Intelligibility is measured by WER for English using Whisper-Large \cite{radford2023robust} and CER for Chinese using Paraformer \cite{gao2022paraformer}.  
Speaker similarity (S-SIM) is computed as the cosine similarity between WavLM-Large speaker embeddings \cite{chen2022wavlm}.  
Transition smoothness is assessed using DNSMOS-Pro (DNSM) \cite{cumlin2024dnsmos} over sliding speech segments, while perceptual quality is evaluated with NISQA \cite{mittag2021nisqa} and OVRL \cite{reddy2022dnsmos}.  
Emotional accuracy is evaluated based on the emotion prompt type, using emotion2vec-Large embeddings \cite{ma2024emotion2vec} for speech prompts and a fine-tuned emotion2vec classifier for text prompts.  
Subjective evaluation is conducted using four MOS criteria: SMOS for speaker similarity, NMOS for the naturalness of emotion transitions, EMOS for emotion alignment, and SPMOS for speaking rate accuracy. All scores are collected on a 5-point scale and reported with mean values and 95\% confidence intervals.

\section{Results and Evaluation}

% ---------------------- Table 2: Duration Scaling ----------------------
\begin{table*}[t]
\centering
\resizebox{\textwidth}{!}{%
\begin{threeparttable}
\begin{tabular}{c l c c c c c c c c}
\toprule
 & \textbf{Model} & \textbf{WER/CER$\downarrow$} & \textbf{DNSM$\uparrow$} &
 \textbf{SSIM$\uparrow$} & \textbf{NISQA$\uparrow$} & \textbf{OVRL$\uparrow$} &
 \textbf{SMOS$\uparrow$} & \textbf{NMOS$\uparrow$} & \textbf{SPMOS$\uparrow$} \\
\midrule

\multicolumn{10}{l}{\textbf{\textit{Speech Emotion Prompt}}} \\
\midrule

% --- English Section ---
\multirow{4}{2.4em}{\centering\rotatebox[origin=c]{90}{English}}
 & MaskGCT & \underline{2.482} & \underline{3.964} & 0.539 & 4.536 & 3.301
 & \underline{4.00}$_{\pm0.32}$ & 3.42$_{\pm0.38}$ & 3.47$_{\pm0.31}$ \\
 & F5TTS & \textbf{1.941} & 3.683 & \underline{0.543} & 4.454 & \underline{3.307}
 & 3.76$_{\pm0.32}$ & 3.02$_{\pm0.42}$ & 3.24$_{\pm0.32}$ \\
 & IndexTTS2 & 2.597 & 3.899 & \textbf{0.575} & \underline{4.604} & 3.273
 & 3.89$_{\pm0.33}$ & \underline{3.87}$_{\pm0.32}$ & \textbf{3.67}$_{\pm0.37}$ \\
 & \cellcolor{gray!15}\textbf{Ours}
 & \cellcolor{gray!15}3.227 & \cellcolor{gray!15}\textbf{3.988} & \cellcolor{gray!15}0.532
 & \cellcolor{gray!15}\textbf{4.766} & \cellcolor{gray!15}\textbf{3.336}
 & \cellcolor{gray!15}\textbf{4.22}$_{\pm0.22}$ & \cellcolor{gray!15}\textbf{4.20}$_{\pm0.25}$ & \cellcolor{gray!15}\underline{3.62}$_{\pm0.34}$ \\

\addlinespace[0.4ex]
\cdashline{1-10}
\addlinespace[0.4ex]

% --- Chinese Section ---
\multirow{4}{2.4em}{\centering\rotatebox[origin=c]{90}{Chinese}}
 & MaskGCT & 8.140 & 3.711 & \textbf{0.614} & \underline{4.366} & 3.167
 & 3.31$_{\pm0.40}$ & 2.60$_{\pm0.39}$ & \underline{3.02}$_{\pm0.31}$ \\
 & F5TTS & 9.004 & 3.386 & \underline{0.598} & 4.286 & 3.204
 & \underline{3.82}$_{\pm0.35}$ & 2.59$_{\pm0.40}$ & 2.89$_{\pm0.36}$ \\
 & IndexTTS2 & \textbf{1.623} & \underline{3.715} & 0.597 & 4.345 & \underline{3.248}
 & 3.76$_{\pm0.34}$ & \underline{3.27}$_{\pm0.34}$ & 2.84$_{\pm0.38}$ \\
 & \cellcolor{gray!15}\textbf{Ours}
 & \cellcolor{gray!15}\underline{2.732} & \cellcolor{gray!15}\textbf{3.803} & \cellcolor{gray!15}0.578
 & \cellcolor{gray!15}\textbf{4.536} & \cellcolor{gray!15}\textbf{3.291}
 & \cellcolor{gray!15}\textbf{3.98}$_{\pm0.28}$ & \cellcolor{gray!15}\textbf{4.16}$_{\pm0.27}$ & \cellcolor{gray!15}\textbf{3.62}$_{\pm0.30}$ \\

\bottomrule
\end{tabular}
\end{threeparttable}
}
\caption{Objective and subjective evaluation on different duration scaling settings. $\downarrow$ indicates that lower values are better, while $\uparrow$ indicates that higher values are better. Subjective results are evaluated by 15 listeners, with 95\% confidence intervals computed using a t-test. The best results are highlighted in \textbf{bold}, and the second-best results are \underline{underlined}.}
\label{com_2}
\end{table*}

\begin{table}[htbp]
\centering
\footnotesize
\resizebox{\linewidth}{!}{
\begin{tabular}{lcccc}
\toprule
\textbf{Method} &
\textbf{WER/CER$\downarrow$} &
\textbf{DNSM$\uparrow$} &
\textbf{SSIM$\uparrow$} &
\textbf{NISQA$\uparrow$} \\
\midrule

\multicolumn{5}{l}{\textbf{\textit{Segment-aware Emotion Conditioning}}} \\
\midrule
\rowcolor{gray!15} \textbf{Ours}                & 2.519 & \textbf{3.925} & \textbf{0.485} & \textbf{4.706} \\
w/o full-text access         & 2.409 & 3.855 & 0.449 & 4.578 \\
w/o alignment                & \textbf{2.043} & 3.831 & 0.442 & 4.639 \\

\midrule
\multicolumn{5}{l}{\textbf{\textit{Segment-aware Duration Steering}}} \\
\midrule
\rowcolor{gray!15} \textbf{Ours} & \textbf{3.227} & \textbf{3.988} & \textbf{0.460} & \textbf{4.766} \\
w/o local steering          & 3.861 & 3.032 & 0.437 & 4.750 \\
w/o global EOS              & 3.513 & 3.885 & 0.451 & 4.717 \\
\bottomrule
\end{tabular}
}
\vspace{-5pt}
\caption{Ablation study of our segment-aware emotion conditioning and duration steering modules.}
\label{abl_1}
\end{table}

\subsection{Comparison with Reference Models}

\noindent\textbf{Objective Evaluation.}
Since comparative methods lack intra-utterance controllability, all segments are synthesized independently and concatenated for evaluation. Under this setting, we conduct objective evaluations for both emotion and duration control. For emotion control, results are reported under two prompting settings: speech emotion prompts and text emotion prompts. For duration control, emotion is fixed to neutral, and segment-level speech synthesis is evaluated under five duration scaling factors (0.75, 0.875, 1.0, 1.125, and 1.25). 
As shown in Tab.~\ref{com_1}, our method achieves the best overall performance on most objective metrics across both languages and prompting settings, with consistent gains on DNSM and SSIM indicating smoother emotion transitions and improved speaker consistency. Although WER/CER and emotion recognition scores are not always optimal, they remain comparable to the IndexTTS2 baseline, which is expected for a training-free framework. For duration control, as shown in Tab.~\ref{com_2}, our method attains the best DNSM, NISQA, and OVRL scores in both languages, reflecting more stable temporal pacing and improved perceptual quality. While some methods achieve higher SSIM, this advantage largely stems from segment-independent synthesis under neutral emotion. In contrast, our method performs multi-segment duration control in a single generation, making SSIM preservation more challenging but better reflecting realistic controllable synthesis scenarios. Overall, these results demonstrate that our training-free framework supports effective intra-utterance emotion and duration control under more challenging settings, while consistently outperforming the baseline and comparative methods on most objective metrics and achieving state-of-the-art transition smoothness.

\noindent\textbf{Subjective Evaluation.}
We report subjective results on SMOS, NMOS, EMOS, and SPMOS in Tab.~\ref{com_1} and \ref{com_2}. Unlike comparative methods that synthesize segments independently, our approach performs one-shot generation with all intra-utterance emotion and duration variations. 
Despite being training-free and inherently bounded by the baseline model, our framework achieves state-of-the-art or highly competitive performance across most MOS metrics in both emotion and duration control evaluations.

\subsection{Ablation Study}

\begin{figure}[htbp]
    \centering
    \includegraphics[width=\columnwidth]{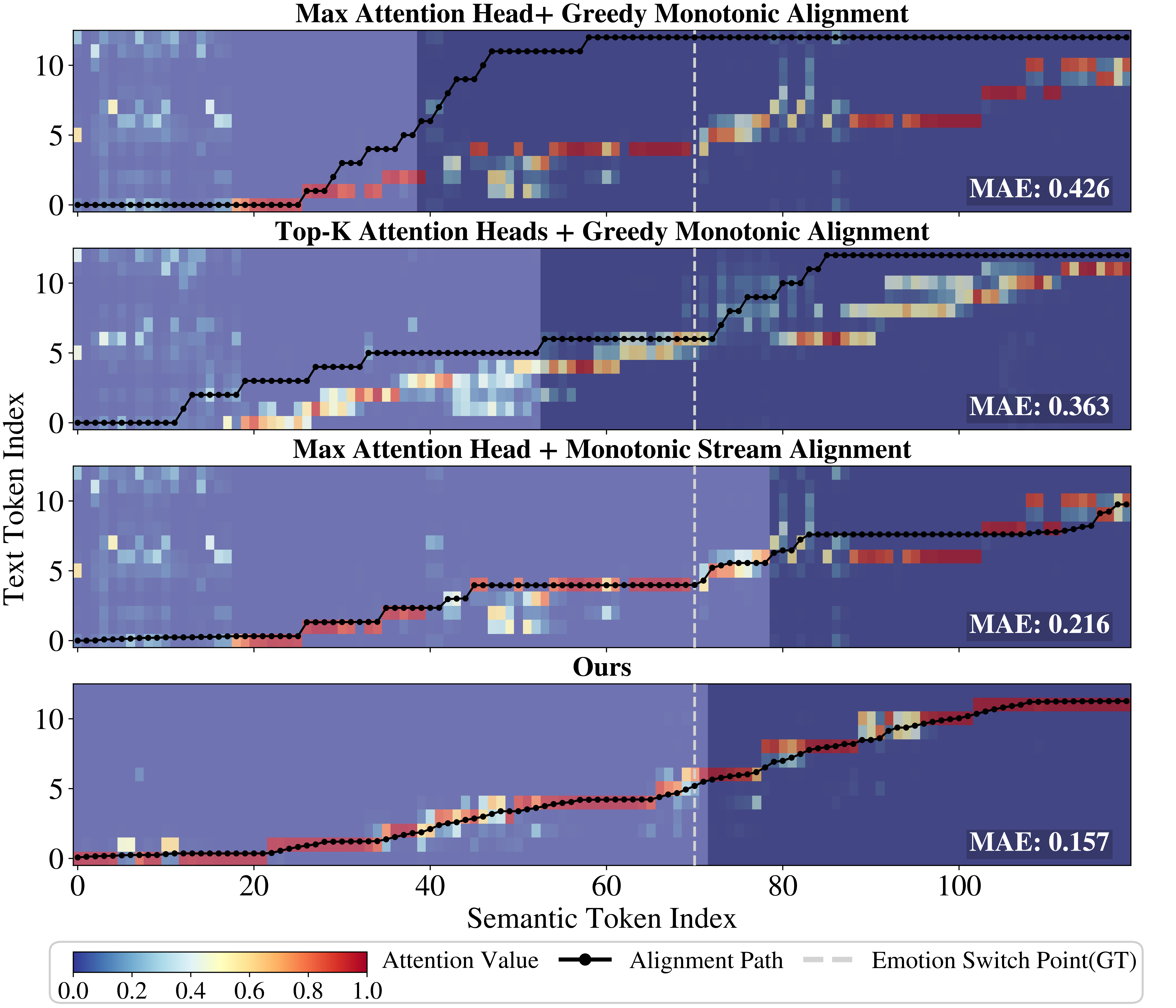}
    \caption{Visualization of alignment paths and emotion switching, with background shading denoting emotion segments and lower MAE indicating better alignment.}
    \label{abl_3}
\end{figure}

\noindent\textbf{Emotion and Duration Control Evaluation.}
We evaluate emotion conditioning and duration steering as two segment-level components of our framework.  
For emotion control, restricting segments to local text only (w/o full-text access) or removing MSA alignment (w/o alignment) degrades expressive quality and cross-segment speaker consistency, as evidenced by reduced DNSM and SSIM in Tab.~\ref{abl_1}, indicating that full-text access and monotonic alignment primarily contribute to smooth emotional transitions rather than token-level accuracy.  
For duration control, disabling local steering (w/o local steering) leads to the largest performance drop, while removing global EOS control (w/o global EOS) causes a smaller but consistent degradation, suggesting that local pacing dominates segment-level naturalness and global EOS provides additional stabilization.

\begin{table}[t!]
\centering
\small
\setlength{\tabcolsep}{1.15mm} 
\begin{tabular}{lccccc}
\toprule
\textbf{Method} & \textbf{$\times$0.75} & \textbf{$\times$0.875} & \textbf{$\times$1} & \textbf{$\times$1.125} & \textbf{$\times$1.25} \\
\midrule
\rowcolor{gray!15} \textbf{Ours} & 3.387 & \textbf{1.704} & \textbf{3.218} & \textbf{3.210} & \textbf{3.211} \\
w/o local steering            & 3.728 & 3.203 & 5.670 & 8.179 & 11.594 \\
w/o global EOS                & \textbf{1.941} & 2.404 & 5.638 & 7.650 & 9.158 \\
Baseline                      & 5.778 & 6.912 & 7.100 & 8.232 & 12.032 \\
\bottomrule
\end{tabular}
\vspace{-6pt}
\caption{Average semantic token number error rate (\%) across segments for duration control under different settings. Lower indicates better duration accuracy.}
\label{abl_2}
\end{table}

\noindent\textbf{Monotonic Stream Alignment Evaluation.}
To evaluate the effectiveness of MSA, we visualize alignment results under different settings in Fig.~\ref{abl_3} and report the mean absolute error (MAE) of segment boundary positions.  
Raw attention maps exhibit diffuse and locally non-monotonic patterns, making greedy alignment highly sensitive to noise and leading to unstable trajectories and frequent segment switching failures.  
Introducing the monotonic stream constraint alleviates this issue and reduces MAE to 0.216, but residual attention uncertainty still causes instability.  
By further incorporating the observation component, where the Top-$k$ most reliable attention heads are selected with $k=3$ instead of relying on a single maximum, MSA effectively suppresses alignment uncertainty, enforces smooth monotonic trajectories, and reduces MAE to 0.157, yielding precise emotion transitions closely aligned with the ground-truth boundaries.

\noindent\textbf{Duration-Specified Evaluation.}
We evaluate duration-specified speech synthesis under five segment-level scaling factors ($\times$0.75, $\times$0.875, $\times$1.0, $\times$1.125, and $\times$1.25), comparing our full system with ablated variants and an IndexTTS2 baseline.  
As shown in Tab.~\ref{abl_2}, our method consistently achieves the lowest semantic token number error across all settings, reducing the error by 3.53\% and 2.41\% on average compared to variants without local steering and global EOS control, respectively. Relative to the baseline without explicit duration control, our approach further yields a 5.07\% average error reduction, demonstrating accurate and robust duration control across diverse segment-level targets.

\begin{figure}[htbp]
    \centering
    \includegraphics[width=\columnwidth]{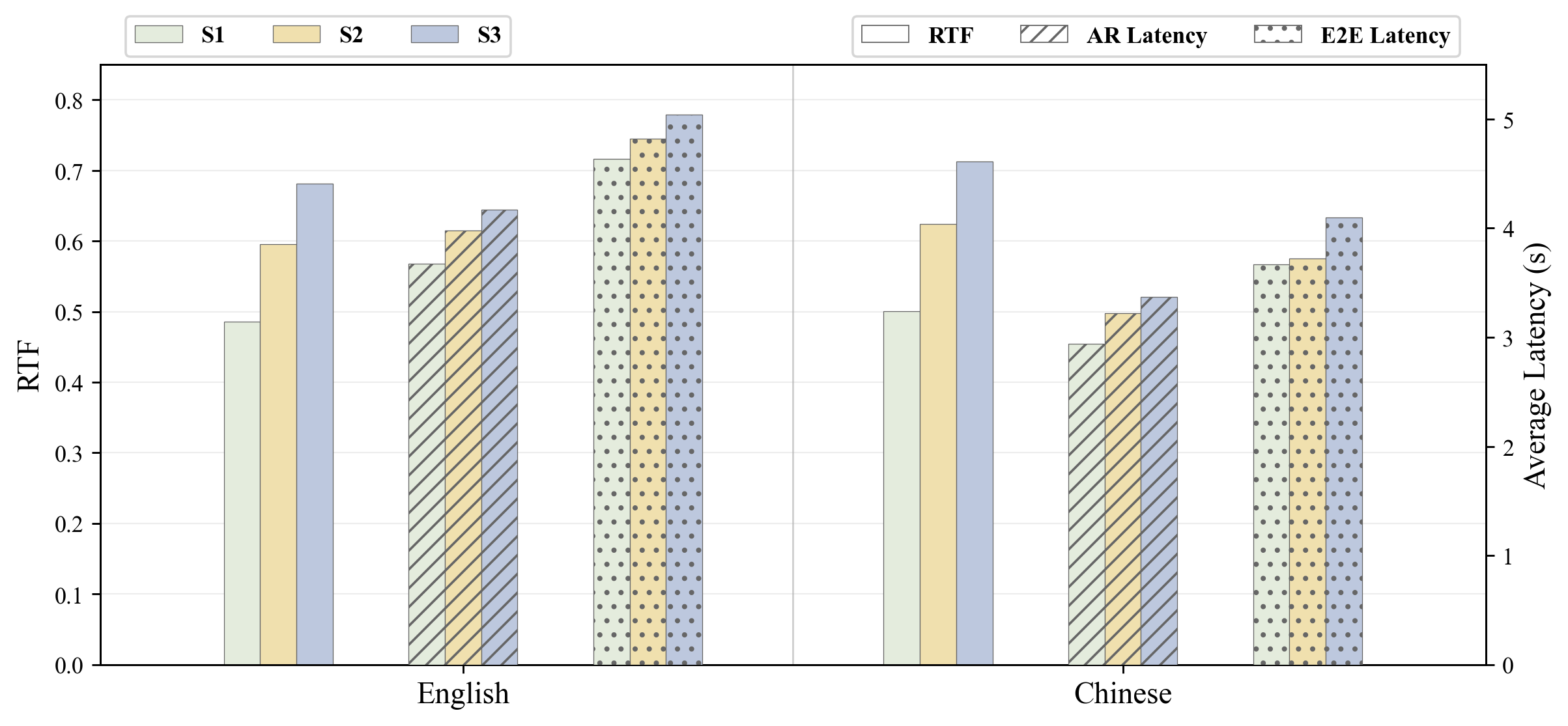}
    \caption{Efficiency analysis of the proposed framework under incremental configurations. S1 denotes the baseline model, S2 adds segment-aware emotion conditioning, and S3 further incorporates segment-aware duration steering. All metrics are evaluated under identical hardware conditions using the NVIDIA RTX 4090 GPU.}
    \label{eff_1}
\end{figure}

\noindent\textbf{Inference Efficiency Evaluation.}
We evaluate the practical efficiency of the proposed framework using real-time factor (RTF), autoregressive (AR) latency, and end-to-end (E2E) latency, where latency is reported as the average inference time in seconds. Starting from the baseline (S1), we progressively incorporate segment-aware emotion conditioning (S2) with 2D causal masking and MSA, followed by segment-aware duration steering (S3) with local duration steering and global EOS steering. As shown in Fig.~\ref{eff_1}, incorporating emotion conditioning results in a moderate increase in computational cost, with RTF rising by 22.7\% on English and 24.5\% on Chinese. Further incorporating duration steering results in additional but controlled overhead, yielding total increases of 40.2\% and 42.2\%, respectively. Despite these increases, RTF remains consistently below 1.0 across both languages, indicating that real-time inference capability is preserved. Meanwhile, latency analysis shows that the majority of the overhead arises from the autoregressive stage. From S1 to S3, AR latency increases by 13.4\% on English and 14.7\% on Chinese, reflecting a favorable trade-off between controllability and efficiency while maintaining real-time performance.

\section{Conclusion}
In this paper, we propose the first training-free controllable framework to enable intra-utterance emotion and duration control in pretrained zero-shot TTS. By introducing segment-aware emotion conditioning and duration steering from an inference-time perspective, our method achieves smooth segment transitions within a single utterance, and is readily applicable to a broad class of autoregressive TTS backbones without requiring architectural modification or additional training. Extensive experiments demonstrate that our method not only delivers state-of-the-art intra-utterance controllability, but also preserves baseline-level speech quality of the underlying TTS model.

\section*{Acknowledgments}
We thank the anonymous reviewers for their valuable comments and suggestions, as well as all participants involved in the subjective evaluations. Qifan Liang, Yuansen Liu, and Ruixin Wei contributed equally to this work. Junchuan Zhao is the corresponding author.

\section*{Limitations}
Despite its advantages, our proposed training-free framework also has several limitations. First, the framework does not explicitly model gradual emotion transitions between adjacent segments. While segment-aware masking and alignment ensure smooth signal-level continuity, emotional variation is controlled in a segment-wise manner rather than through a continuous emotion trajectory, which may limit the representation of intermediate emotional states. Second, the precision of duration control is influenced by the duration representation learned in the pretrained baseline TTS model. Since our approach operates without parameter updates, the duration embedding may not always support strictly linear or fine-grained timing control, particularly under highly expressive or out-of-domain conditions. Future work will investigate training-free or minimally adaptive strategies to better model continuous emotion evolution and duration precision, while preserving the simplicity and generality of the proposed framework.

\section*{Ethical Considerations}
This work involves the use of large language models to generate a synthetic text dataset for Qwen3 fine-tuning and model evaluation, and therefore shares some general characteristics of LLM-based generation, such as occasional variations in factual precision or stylistic expression. All models and datasets used are publicly available and employed under their respective licenses, and no private or personally identifiable speech data is involved. While intra-utterance-level controllable TTS can benefit expressive speech synthesis and human-computer interaction research, high-fidelity speech generation also entails potential risks if misused, such as speaker impersonation or spoofing of voice-based authentication systems. In practical applications, it is important to incorporate appropriate safeguards, including audio watermarking, output traceability, or dedicated detection models, to facilitate the identification of synthesized speech and discourage unintended or malicious misuse.

\bibliography{main}

@inproceedings{eskimez2024e2,
  author       = {Sefik Emre Eskimez and
                  Xiaofei Wang and
                  Manthan Thakker and
                  Canrun Li and
                  Chung{-}Hsien Tsai and
                  Zhen Xiao and
                  Hemin Yang and
                  Zirun Zhu and
                  Min Tang and
                  Xu Tan and
                  Yanqing Liu and
                  Sheng Zhao and
                  Naoyuki Kanda},
  title        = {{E2} {TTS:} Embarrassingly Easy Fully Non-Autoregressive Zero-Shot
                  {TTS}},
  booktitle    = {{IEEE} Spoken Language Technology Workshop, {SLT} 2024},
  pages        = {682--689},
  publisher    = {{IEEE}},
  year         = {2024},
  doi          = {10.1109/SLT61566.2024.10832320},
}

@article{du2024cosyvoice1,
  title={Cosyvoice: A scalable multilingual zero-shot text-to-speech synthesizer based on supervised semantic tokens},
  author={Du, Zhihao and Chen, Qian and Zhang, Shiliang and Hu, Kai and Lu, Heng and Yang, Yexin and Hu, Hangrui and Zheng, Siqi and Gu, Yue and Ma, Ziyang and et al.},
  journal={arXiv preprint arXiv:2407.05407},
  year={2024}
}

@inproceedings{wang2025maskgct,
  author       = {Yuancheng Wang and
                  Haoyue Zhan and
                  Liwei Liu and
                  Ruihong Zeng and
                  Haotian Guo and
                  Jiachen Zheng and
                  Qiang Zhang and
                  Xueyao Zhang and
                  Shunsi Zhang and
                  Zhizheng Wu},
  title        = {MaskGCT: Zero-Shot Text-to-Speech with Masked Generative Codec Transformer},
  booktitle    = {The Thirteenth International Conference on Learning Representations,
                  {ICLR} 2025},
  year         = {2025},
}

@article{wang2025spark,
  title={Spark-tts: An efficient llm-based text-to-speech model with single-stream decoupled speech tokens},
  author={Wang, Xinsheng and Jiang, Mingqi and Ma, Ziyang and Zhang, Ziyu and Liu, Songxiang and Li, Linqin and Liang, Zheng and Zheng, Qixi and Wang, Rui and Feng, Xiaoqin},
  journal={arXiv preprint arXiv:2503.01710},
  year={2025}
}

@inproceedings{chen2025f5,
  author       = {Yushen Chen and
                  Zhikang Niu and
                  Ziyang Ma and
                  Keqi Deng and
                  Chunhui Wang and
                  Jian Zhao and
                  Kai Yu and
                  Xie Chen},
  title        = {{F5-TTS:} {A} Fairytaler that Fakes Fluent and Faithful Speech with Flow Matching},
  booktitle    = {Proceedings of the 63rd Annual Meeting of the Association for Computational Linguistics (Volume 1: Long Papers), {ACL} 2025},
  pages        = {6255--6271},
  publisher    = {Association for Computational Linguistics},
  year         = {2025},
}

@inproceedings{guo2023emodiff,
  author       = {Yiwei Guo and
                  Chenpeng Du and
                  Xie Chen and
                  Kai Yu},
  title        = {Emodiff: Intensity Controllable Emotional Text-to-Speech with Soft-Label Guidance},
  booktitle    = {{IEEE} International Conference on Acoustics, Speech and Signal Processing {ICASSP} 2023},
  pages        = {1--5},
  publisher    = {{IEEE}},
  year         = {2023},
  doi          = {10.1109/ICASSP49357.2023.10095621},
}

@inproceedings{kang2023zet,
  author       = {Minki Kang and
                  Wooseok Han and
                  Sung Ju Hwang and
                  Eunho Yang},
  title        = {ZET-Speech: Zero-shot adaptive Emotion-controllable Text-to-Speech Synthesis with Diffusion and Style-based Models},
  booktitle    = {24th Annual Conference of the International Speech Communication Association, Interspeech 2023},
  pages        = {4339--4343},
  publisher    = {{ISCA}},
  year         = {2023},
  doi          = {10.21437/INTERSPEECH.2023-754},
}

@inproceedings{diatlova2023emospeech,
  author       = {Daria Diatlova and
                  Vitalii Shutov},
  title        = {EmoSpeech: guiding FastSpeech2 towards Emotional Text to Speech},
  booktitle    = {12th {ISCA} Speech Synthesis Workshop, {SSW} 2023},
  pages        = {106--112},
  publisher    = {{ISCA}},
  year         = {2023},
  doi          = {10.21437/SSW.2023-17}
}

@inproceedings{tang2024ed,
  author       = {Haobin Tang and
                  Xulong Zhang and
                  Ning Cheng and
                  Jing Xiao and
                  Jianzong Wang},
  title        = {{ED-TTS:} Multi-Scale Emotion Modeling Using Cross-Domain Emotion Diarization for Emotional Speech Synthesis},
  booktitle    = {{IEEE} International Conference on Acoustics, Speech and Signal Processing, {ICASSP} 2024},
  pages        = {12146--12150},
  publisher    = {{IEEE}},
  year         = {2024},
  doi          = {10.1109/ICASSP48485.2024.10446467}
}

@inproceedings{gao2025emo,
  author       = {Xiaoxue Gao and
                  Chen Zhang and
                  Yiming Chen and
                  Huayun Zhang and
                  Nancy F. Chen},
  title        = {Emo-DPO: Controllable Emotional Speech Synthesis through Direct Preference Optimization},
  booktitle    = {2025 {IEEE} International Conference on Acoustics, Speech and Signal Processing, {ICASSP} 2025},
  pages        = {1--5},
  publisher    = {{IEEE}},
  year         = {2025},
  doi          = {10.1109/ICASSP49660.2025.10888737}
}

@inproceedings{guo2023prompttts,
  author       = {Zhifang Guo and
                  Yichong Leng and
                  Yihan Wu and
                  Sheng Zhao and
                  Xu Tan},
  title        = {Prompttts: Controllable Text-To-Speech With Text Descriptions},
  booktitle    = {{IEEE} International Conference on Acoustics, Speech and Signal Processing {ICASSP} 2023},
  pages        = {1--5},
  publisher    = {{IEEE}},
  year         = {2023},
  doi          = {10.1109/ICASSP49357.2023.10096285},
}

@inproceedings{liu2023promptstyle,
  author       = {Guanghou Liu and
                  Yongmao Zhang and
                  Yi Lei and
                  Yunlin Chen and
                  Rui Wang and
                  Lei Xie and
                  Zhifei Li},
  title        = {PromptStyle: Controllable Style Transfer for Text-to-Speech with Natural Language Descriptions},
  booktitle    = {24th Annual Conference of the International Speech Communication Association, Interspeech 2023},
  pages        = {4888--4892},
  publisher    = {{ISCA}},
  year         = {2023},
  doi          = {10.21437/INTERSPEECH.2023-1779}
}

@article{yang2024instructtts,
  author       = {Dongchao Yang and
                  Songxiang Liu and
                  Rongjie Huang and
                  Chao Weng and
                  Helen Meng},
  title        = {InstructTTS: Modelling Expressive {TTS} in Discrete Latent Space With Natural Language Style Prompt},
  journal      = {{IEEE} {ACM} Trans. Audio Speech Lang. Process.},
  volume       = {32},
  pages        = {2913--2925},
  year         = {2024},
  doi          = {10.1109/TASLP.2024.3402088}
}

@inproceedings{yang2025emovoice,
author = {Yang, Guanrou and Yang, Chen and Chen, Qian and Ma, Ziyang and Chen, Wenxi and Wang, Wen and Wang, Tianrui and Yang, Yifan and Niu, Zhikang and Liu, Wenrui and Yu, Fan and Du, Zhihao and Gao, Zhifu and Zhang, Shiliang and Chen, Xie},
title = {EmoVoice: LLM-based Emotional Text-To-Speech Model with Freestyle Text Prompting},
year = {2025},
publisher = {{ACM}},
doi = {10.1145/3746027.3754829},
booktitle = {Proceedings of the 33rd ACM International Conference on Multimedia},
pages = {10748–10757},
numpages = {10}
}

@article{du2024cosyvoice2,
  title={Cosyvoice 2: Scalable streaming speech synthesis with large language models},
  author={Du, Zhihao and Wang, Yuxuan and Chen, Qian and Shi, Xian and Lv, Xiang and Zhao, Tianyu and Gao, Zhifu and Yang, Yexin and Gao, Changfeng and Wang, Hui and et al.},
  journal={arXiv preprint arXiv:2412.10117},
  year={2024}
}

@article{zhou2025indextts2,
  title={IndexTTS2: A Breakthrough in Emotionally Expressive and Duration-Controlled Auto-Regressive Zero-Shot Text-to-Speech},
  author={Zhou, Siyi and Zhou, Yiquan and He, Yi and Zhou, Xun and Wang, Jinchao and Deng, Wei and Shu, Jingchen},
  journal={arXiv preprint arXiv:2506.21619},
  year={2025}
}

@article{tan2024naturalspeech,
  author       = {Xu Tan and
                  Jiawei Chen and
                  Haohe Liu and
                  Jian Cong and
                  Chen Zhang and
                  Yanqing Liu and
                  Xi Wang and
                  Yichong Leng and
                  Yuanhao Yi and
                  Lei He and
                  Sheng Zhao and
                  Tao Qin and
                  Frank K. Soong and
                  Tie{-}Yan Liu},
  title        = {NaturalSpeech: End-to-End Text-to-Speech Synthesis With Human-Level Quality},
  journal      = {{IEEE} Trans. Pattern Anal. Mach. Intell.},
  volume       = {46},
  number       = {6},
  pages        = {4234--4245},
  year         = {2024},
  doi          = {10.1109/TPAMI.2024.3356232},
}

@inproceedings{luo2024emotion,
  author       = {Xuan Luo and
                  Shinnosuke Takamichi and
                  Tomoki Koriyama and
                  Yuki Saito and
                  Hiroshi Saruwatari},
  title        = {Emotion-Controllable Speech Synthesis Using Emotion Soft Labels and Fine-Grained Prosody Factors},
  booktitle    = {Asia-Pacific Signal and Information Processing Association Annual Summit and Conference, {APSIPA} {ASC} 2021},
  pages        = {794--799},
  publisher    = {{IEEE}},
  year         = {2021},
}

@inproceedings{im2022emoq,
  author       = {Chae{-}Bin Im and
                  Sang{-}Hoon Lee and
                  Seung{-}Bin Kim and
                  Seong{-}Whan Lee},
  title        = {{EMOQ-TTS:} Emotion Intensity Quantization for Fine-Grained Controllable Emotional Text-to-Speech},
  booktitle    = {{IEEE} International Conference on Acoustics, Speech and Signal Processing, {ICASSP} 2022},
  pages        = {6317--6321},
  publisher    = {{IEEE}},
  year         = {2022},
  doi          = {10.1109/ICASSP43922.2022.9747098},
}

@article{kanda2024making,
  title={Making flow-matching-based zero-shot text-to-speech laugh as you like},
  author={Kanda, Naoyuki and Wang, Xiaofei and Eskimez, Sefik Emre and Thakker, Manthan and Yang, Hemin and Zhu, Zirun and Tang, Min and Li, Canrun and Tsai, Chung-Hsien and Xiao, Zhen},
  journal={arXiv preprint arXiv:2402.07383},
  year={2024}
}

@inproceedings{wu2024laugh,
  author       = {Haibin Wu and
                  Xiaofei Wang and
                  Sefik Emre Eskimez and
                  Manthan Thakker and
                  Daniel Tompkins and
                  Chung{-}Hsien Tsai and
                  Canrun Li and
                  Zhen Xiao and
                  Sheng Zhao and
                  Jinyu Li and
                  Naoyuki Kanda},
  title        = {Laugh Now Cry Later: Controlling Time-Varying Emotional States of Flow-Matching-Based Zero-Shot Text-To-Speech},
  booktitle    = {{IEEE} Spoken Language Technology Workshop, {SLT} 2024},
  pages        = {690--697},
  publisher    = {{IEEE}},
  year         = {2024},
  doi          = {10.1109/SLT61566.2024.10832181},
}

@article{wang2025word,
  title={Word-Level Emotional Expression Control in Zero-Shot Text-to-Speech Synthesis},
  author={Wang, Tianrui and Wang, Haoyu and Ge, Meng and Gong, Cheng and Qiang, Chunyu and Ma, Ziyang and Huang, Zikang and Yang, Guanrou and Wang, Xiaobao and Chng, Eng Siong},
  journal={arXiv preprint arXiv:2509.24629},
  year={2025}
}

@inproceedings{lee2025dittotts,
  author       = {Keon Lee and
                  Dong Won Kim and
                  Jaehyeon Kim and
                  Seungjun Chung and
                  Jaewoong Cho},
  title        = {DiTTo-TTS: Diffusion Transformers for Scalable Text-to-Speech without Domain-Specific Factors},
  booktitle    = {The Thirteenth International Conference on Learning Representations, {ICLR} 2025},
  year         = {2025},
}

@article{kim2023sc,
  title={SC VALL-E: Style-controllable zero-shot text to speech synthesizer},
  author={Kim, Daegyeom and Hong, Seongho and Choi, Yong-Hoon},
  journal={arXiv preprint arXiv:2307.10550},
  year={2023}
}

@inproceedings{zhou2024voxinstruct,
  author       = {Yixuan Zhou and
                  Xiaoyu Qin and
                  Zeyu Jin and
                  Shuoyi Zhou and
                  Shun Lei and
                  Songtao Zhou and
                  Zhiyong Wu and
                  Jia Jia},
  title        = {VoxInstruct: Expressive Human Instruction-to-Speech Generation with Unified Multilingual Codec Language Modelling},
  booktitle    = {Proceedings of the 32nd {ACM} International Conference on Multimedia, {MM} 2024},
  pages        = {554--563},
  publisher    = {{ACM}},
  year         = {2024},
  doi          = {10.1145/3664647.3681680},
}

@article{li2025flespeech,
  title={Flespeech: Flexibly controllable speech generation with various prompts},
  author={Li, Hanzhao and Li, Yuke and Wang, Xinsheng and Hu, Jingbin and Xie, Qicong and Yang, Shan and Xie, Lei},
  journal={arXiv preprint arXiv:2501.04644},
  year={2025}
}

@article{du2025cosyvoice3,
  title={Cosyvoice 3: Towards in-the-wild speech generation via scaling-up and post-training},
  author={Du, Zhihao and Gao, Changfeng and Wang, Yuxuan and Yu, Fan and Zhao, Tianyu and Wang, Hao and Lv, Xiang and Wang, Hui and Ni, Chongjia and Shi, Xian and et al.},
  journal={arXiv preprint arXiv:2505.17589},
  year={2025}
}

@inproceedings{sahipjohn2024dubwise,
  author       = {Neha Sahipjohn and
                  Ashishkumar Gudmalwar and
                  Nirmesh Shah and
                  Pankaj Wasnik and
                  Rajiv Ratn Shah},
  title        = {DubWise: Video-Guided Speech Duration Control in Multimodal LLM-based Text-to-Speech for Dubbing},
  booktitle    = {25th Annual Conference of the International Speech Communication Association, Interspeech 2024},
  publisher    = {{ISCA}},
  year         = {2024},
  doi          = {10.21437/INTERSPEECH.2024-1700},
}

@inproceedings{chen-etal-2024-emoknob,
  author       = {Haozhe Chen and
                  Run Chen and
                  Julia Hirschberg},
  editor       = {Yaser Al{-}Onaizan and
                  Mohit Bansal and
                  Yun{-}Nung Chen},
  title        = {EmoKnob: Enhance Voice Cloning with Fine-Grained Emotion Control},
  booktitle    = {Proceedings of the 2024 Conference on Empirical Methods in Natural Language Processing, {EMNLP} 2024},
  pages        = {8170--8180},
  publisher    = {Association for Computational Linguistics},
  year         = {2024},
  doi          = {10.18653/V1/2024.EMNLP-MAIN.466},
}

@article{lam2025present,
  author       = {Perry Lam and
                  Huayun Zhang and
                  Nancy F. Chen and
                  Berrak Sisman and
                  Dorien Herremans},
  title        = {{PRESENT:} Zero-Shot Text-to-Prosody Control},
  journal      = {{IEEE} Signal Process. Lett.},
  volume       = {32},
  pages        = {776--780},
  year         = {2025},
  doi          = {10.1109/LSP.2025.3528359},
}

@inproceedings{suni2025style,
  title     = {{Style and Prosody control for Zero-shot Speech Synthesis}},
  author    = {Antti Suni and Sébastien {Le Maguer} and Sofoklis Kakouros and Tuukka Törö and Juraj Šimko},
  year      = {2025},
  booktitle = {{13th edition of the Speech Synthesis Workshop}},
  pages     = {28--34},
  doi       = {10.21437/SSW.2025-5},
}

@article{xie2025emosteer,
  title={EmoSteer-TTS: Fine-Grained and Training-Free Emotion-Controllable Text-to-Speech via Activation Steering},
  author={Xie, Tianxin and Yang, Shan and Li, Chenxing and Yu, Dong and Liu, Li},
  journal={arXiv preprint arXiv:2508.03543},
  year={2025}
}

@article{zhou2022emotional,
  author       = {Kun Zhou and
                  Berrak Sisman and
                  Rui Liu and
                  Haizhou Li},
  title        = {Emotional voice conversion: Theory, databases and {ESD}},
  journal      = {Speech Commun.},
  volume       = {137},
  pages        = {1--18},
  year         = {2022},
  doi          = {10.1016/J.SPECOM.2021.11.006},
}

@inproceedings{radford2023robust,
  author       = {Alec Radford and
                  Jong Wook Kim and
                  Tao Xu and
                  Greg Brockman and
                  Christine McLeavey and
                  Ilya Sutskever},
  title        = {Robust Speech Recognition via Large-Scale Weak Supervision},
  booktitle    = {International Conference on Machine Learning, {ICML} 2023},
  volume       = {202},
  pages        = {28492--28518},
  publisher    = {{PMLR}},
  year         = {2023},
}

@inproceedings{gao2022paraformer,
  author       = {Zhifu Gao and
                  Shiliang Zhang and
                  Ian McLoughlin and
                  Zhijie Yan},
  title        = {Paraformer: Fast and Accurate Parallel Transformer for Non-autoregressive End-to-End Speech Recognition},
  booktitle    = {23rd Annual Conference of the International Speech Communication Association,
                  Interspeech 2022},
  pages        = {2063--2067},
  publisher    = {{ISCA}},
  year         = {2022},
  doi          = {10.21437/INTERSPEECH.2022-9996},
}

@article{chen2022wavlm,
  author       = {Sanyuan Chen and
                  Chengyi Wang and
                  Zhengyang Chen and
                  Yu Wu and
                  Shujie Liu and
                  Zhuo Chen and
                  Jinyu Li and
                  Naoyuki Kanda and
                  Takuya Yoshioka and
                  Xiong Xiao and
                  Jian Wu and
                  Long Zhou and
                  Shuo Ren and
                  Yanmin Qian and
                  Yao Qian and
                  Jian Wu and
                  Michael Zeng and
                  Xiangzhan Yu and
                  Furu Wei},
  title        = {WavLM: Large-Scale Self-Supervised Pre-Training for Full Stack Speech Processing},
  journal      = {{IEEE} J. Sel. Top. Signal Process.},
  volume       = {16},
  number       = {6},
  pages        = {1505--1518},
  year         = {2022},
  doi          = {10.1109/JSTSP.2022.3188113},
}

@inproceedings{cumlin2024dnsmos,
  author       = {Fredrik Cumlin and
                  Xinyu Liang and
                  Victor Ungureanu and
                  Chandan K. A. Reddy and
                  Christian Sch{\"{u}}ldt and
                  Saikat Chatterjee},
  title        = {{DNSMOS} Pro: {A} Reduced-Size {DNN} for Probabilistic {MOS} of Speech},
  booktitle    = {25th Annual Conference of the International Speech Communication Association, Interspeech 2024},
  publisher    = {{ISCA}},
  year         = {2024},
  doi          = {10.21437/INTERSPEECH.2024-478},
}

@inproceedings{mittag2021nisqa,
  author       = {Gabriel Mittag and
                  Babak Naderi and
                  Assmaa Chehadi and
                  Sebastian M{\"{o}}ller},
  title        = {{NISQA:} {A} Deep CNN-Self-Attention Model for Multidimensional Speech Quality Prediction with Crowdsourced Datasets},
  booktitle    = {22nd Annual Conference of the International Speech Communication Association, Interspeech 2021},
  pages        = {2127--2131},
  publisher    = {{ISCA}},
  year         = {2021},
  doi          = {10.21437/INTERSPEECH.2021-299},
}

@inproceedings{ma2024emotion2vec,
  author       = {Ziyang Ma and
                  Zhisheng Zheng and
                  Jiaxin Ye and
                  Jinchao Li and
                  Zhifu Gao and
                  Shiliang Zhang and
                  Xie Chen},
  title        = {emotion2vec: Self-Supervised Pre-Training for Speech Emotion Representation},
  booktitle    = {Findings of the Association for Computational Linguistics, {ACL} 2024},
  pages        = {15747--15760},
  publisher    = {Association for Computational Linguistics},
  year         = {2024},
  doi          = {10.18653/V1/2024.FINDINGS-ACL.931},
}

@inproceedings{reddy2022dnsmos,
  author       = {Chandan K. A. Reddy and
                  Vishak Gopal and
                  Ross Cutler},
  title        = {Dnsmos {P.835:} {A} Non-Intrusive Perceptual Objective Speech Quality Metric to Evaluate Noise Suppressors},
  booktitle    = {{IEEE} International Conference on Acoustics, Speech and Signal Processing, {ICASSP} 2022},
  pages        = {886--890},
  publisher    = {{IEEE}},
  year         = {2022},
  doi          = {10.1109/ICASSP43922.2022.9746108},
}

@inproceedings{tang2023squad,
  title={SQuAD-SRC: A Dataset for Multi-Accent Spoken Reading Comprehension.},
  author={Tang, Yixuan and Tung, Anthony KH and Elkind, Edith},
  booktitle={IJCAI},
  pages={5206--5214},
  year={2023}
}

@article{scherer2003vocal,
  title={Vocal communication of emotion: A review of research paradigms},
  author={Scherer, Klaus R},
  journal={Speech communication},
  volume={40},
  number={1-2},
  pages={227--256},
  year={2003},
  publisher={Elsevier}
}

% \clearpage
\appendix

\section{Automatic Prompt Construction}
\label{appendix:A}
\subsection{LLM-Based Prompting Strategy}
MED-TTS is synthesized through a structured LLM-driven pipeline consisting of content generation, segment-level annotation, and manual verification. Specifically, we adopt a two-stage prompting strategy to automatically construct intra-utterance emotion and duration TTS prompts. In Step~1, GPT-4o\footnote{\url{https://openai.com/index/hello-gpt-4o/}} is prompted to generate 15,000 English and 15,000 Chinese emotion-rich content texts that explicitly exhibit continuous emotional transitions within a single utterance. Each text spans multiple emotional phases drawn from 7 core emotions (\emph{happy, sad, angry, surprised, fearful, disgusted,} and \emph{neutral}), forming either smooth or abrupt emotional progressions rather than a single static emotional state. In Step~2, DeepSeek-Chat \footnote{\url{https://api-docs.deepseek.com/}} is prompted to leverage its strong contextual understanding capability, performing precise semantic decomposition by segmenting the utterance into multiple emotion-specific segments. For each segment, it produces a concise natural language emotion description together with an estimated speaking duration, forming a structured sequence of emotion-duration pairs that directly aligns with the input requirements of controllable TTS systems. Example prompts used in Step~1 and Step~2 are shown in List.~\ref{lst1} and List.~\ref{lst2}, respectively.

\begin{table*}[t!]
\centering
\resizebox{0.95\textwidth}{!}{%
\begin{threeparttable}

\begin{tabular}{c l c c m{12em} c m{12em}}
\toprule
\textbf{Language} & 
\multicolumn{1}{c}{\textbf{Category}} & 
\textbf{Count} & 
\textbf{Duration(h)} & 
\multicolumn{1}{c}{\textbf{Text Example}} & 
\textbf{Emotion Sequence Example} & 
\multicolumn{1}{c}{\textbf{Emotion Description Example}} \\
\midrule

% ========================== CHINESE ==========================
% span 3 rows (lines)
\multirow{3}{*}[-4em]{Chinese} & 
Emotional Dialogue & 4,996 & 9.89 & 
\cn{失去你的日子里，心中满是空虚。}$\Rightarrow$ \cn{但与你重逢的那一刻，我的笑容重新绽放。} & 
Sad $\Rightarrow$ Happy & 
\cn{语速缓慢，语调低沉，带有失落和空虚感。}$\Rightarrow$ \cn{语速轻快，语调上扬，充满喜悦和温暖。} \\
\cmidrule{2-7}

& 
Observational Phrase & 4,989 & 10.00 & 
\cn{茶杯中水波平静，}$\Rightarrow$ \cn{内心却如火山爆发。}& 
Neutral $\Rightarrow$ Angry & 
\cn{语调平稳，语速适中，声音自然放松。}$\Rightarrow$ \cn{语速加快，音调升高，声音紧张有力。} \\
\cmidrule{2-7}

& 
Vivid Description & 4,980 & 10.05 & 
\cn{她无意中推开暗门，}$\Rightarrow$ \cn{霉味扑鼻，}$\Rightarrow$ \cn{脚步却不敢移动。} & 
Surprised $\Rightarrow$ Disgusted $\Rightarrow$ Fearful & 
\cn{语调突然上扬，语速稍快，带有意外感。}$\Rightarrow$ \cn{声音压低，语速放缓，带有明显的嫌恶和停顿。}$\Rightarrow$ \cn{语调紧张、迟疑，语速缓慢，伴有轻微颤抖。} \\

\midrule

% ========================== ENGLISH ==========================
\multirow{3}{*}[-6em]{English} & 
Emotional Dialogue & 5,034 & 9.79 & 
What in the world is that? $\Rightarrow$ Ugh, it's revolting. $\Rightarrow$ Well, I suppose it's just another part of life.& 
Surprised $\Rightarrow$ Disgusted $\Rightarrow$ Neutral & 
Voice rises sharply in pitch, with a quick, breathy delivery. $\Rightarrow$ Tone is low, guttural, and drawn out with a visceral recoil. $\Rightarrow$ Pace evens out to a calm, steady, and slightly resigned rhythm. \\
\cmidrule{2-7}

& 
Observational Phrase & 5,029 & 9.56 & 
A heated debate burned fiercely, each word adding fuel, $\Rightarrow$ until playful banter extinguished the flames with lighthearted ease. & 
Angry $\Rightarrow$ Happy & 
Voice is sharp, intense, and rapid, with a clipped, aggressive edge. $\Rightarrow$ Tone becomes warm, relaxed, and lilting, with a cheerful, flowing cadence. \\
\cmidrule{2-7}

& 
Vivid Description & 5,029 & 9.85 & 
A high-pitched scream pierced his thoughts, $\Rightarrow$ unraveling into a soft sigh, weighted with heartache and longing. & 
Fearful $\Rightarrow$ Sad & 
Voice is sharp, tense, and sudden, with a quick, breathy delivery. $\Rightarrow$ Tone is slow, breathy, and heavy, with a drawn-out, mournful quality. \\

\midrule
% ========================== TOTAL ROW ==========================
\textbf{Total} & & \textbf{30,057} & \textbf{59.14} & \multicolumn{3}{c}{} \\

\bottomrule
\end{tabular}
\end{threeparttable}
}
\caption{Dataset statistics and representative examples across languages and text categories.}
\label{tab:phase2_stats_centered}
\end{table*}

\subsection{Quality Control and Manual Verification}

To ensure the reliability of the constructed dataset, we employ a combination of automated validation and manual verification for both Step~1 text generation and Step~2 multi-segment prompt annotation.

At the automated level, we employ validation scripts to enforce strict structural, semantic, and distributional constraints. For Step~1 text generation, each sample is verified for valid JSON formatting and required fields, including the text content, text category, and emotion sequence. Specifically, text length is constrained to 25-45 words for English and 20-30 characters for Chinese texts, and each emotion sequence is restricted to 2-3 segments drawn from a predefined set of 7 emotion categories. To reduce redundancy, we remove exact duplicates and filter near-duplicate texts using similarity-based criteria computed from normalized sequence-matching scores between token sequences, applying an overall similarity threshold of 0.85 and an opening similarity threshold of 0.5 over the first few tokens, with similarity comparisons primarily performed among samples sharing the same emotion sequence. For Step~2 multi-segment prompt annotation, automated checks enforce strict alignment between the text and its associated emotion annotations. Specifically, the number and order of segments are required to exactly match the predefined emotion sequence. Each segment must include a valid emotion label, a non-empty natural language emotion description with constrained length, and an estimated speaking duration falling within a predefined range of 0.3-8.0 seconds.

Beyond automated filtering, we conduct manual verification through stratified random sampling of the validated outputs. In total, 1,000 samples are randomly selected for human review, comprising 500 English and 500 Chinese samples. The sampling process is stratified by language, text category, and the number of emotion segments to ensure broad coverage across diverse data conditions. Human reviewers then examine segmentation boundaries, emotion-text alignment, vocal-affect descriptions, and the plausibility of estimated speaking durations to identify subtle issues that may escape rule-based automatic checks. Insights obtained from this process are used to iteratively refine prompting strategies and validation thresholds. The manual verification checklist applied in Step~1 and Step~2 is provided in List~\ref{lst3}.

\subsection{Dataset Statistics and Distribution}

We summarize the statistics and distribution of the MED-TTS dataset across languages, text categories, and emotion segments. As illustrated in Fig.~\ref{dataset_method:a}, the dataset is well balanced across languages, comprising 14,965 Chinese and 15,092 English samples. Within each language, samples are further evenly distributed across three text categories (vivid descriptions, emotional dialogues, and observational phrases), each contributing approximately 5,000 utterances. A finer-grained breakdown reveals that, within every text category, utterances containing three emotion segments consistently outnumber those with two emotion segments (e.g., roughly 4,100 vs. 800 per category), reflecting a deliberate emphasis on richer intra-utterance emotional transitions. Fig.~\ref{dataset_method:b} presents the segment-level emotion statistics. Across both Chinese and English, the 7 emotion types are uniformly represented, with each emotion accounting for approximately 1,200 segments. The average segment length remains stable within each language but differs across languages, with Chinese emotional segments typically spanning about 24-26 characters, while English segments are longer on average, ranging from roughly 34-39 words depending on emotion. Overall, MED-TTS achieves structured balance across languages, text categories, and emotion types, while maintaining sufficient emphasis on multi-emotion utterances to support modeling of continuous intra-utterance transitions.

As illustrated in Tab.~\ref{tab:phase2_stats_centered}, we further provide representative examples from the MED-TTS dataset across the three text categories for both Chinese and English. For each language-category pair, the table reports the sample count and total duration, along with illustrative text examples, corresponding emotion sequences, and natural language emotion descriptions. These examples demonstrate that each category consistently includes high-quality samples with different numbers of emotion segments, highlighting the dataset’s coverage of diverse content types and intra-utterance emotional structures across different languages.

\subsection{Fine-tuning Details}

To enable automatic construction of segment-level TTS prompts, we fine-tune the Qwen3-8B large language model via supervised instruction tuning with parameter-efficient adaptation. We adopt LoRA to update only low-rank adapters while keeping the backbone frozen, thereby preserving general linguistic capabilities. Specifically, fine-tuning is carried out using the SFT-Trainer framework, with LoRA adapters applied to the attention and feed-forward projection layers, using a rank of 32, a scaling factor of 64, and a dropout rate of 0.1. Training is performed for 4 epochs with a per-device batch size of 2 and gradient accumulation over 4 steps, yielding an effective batch size of 8. We use a learning rate of $1\times10^{-4}$ with a linear warmup of 100 steps and enable mixed-precision FP16 training for efficiency. 

\section{Segment-Aware Emotion Conditioning}
\label{appendix:B}
\begin{figure*}[t]
  \centering
  \begin{subfigure}{0.22\textwidth}
    \centering
    \includegraphics[width=\linewidth]{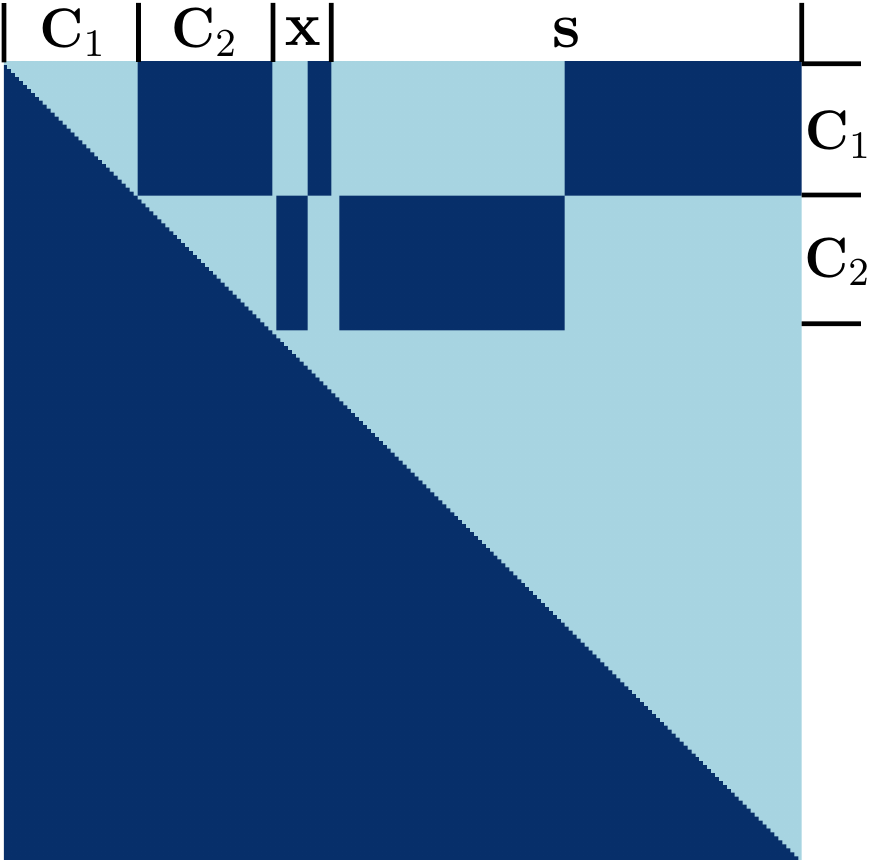}
    \caption{$M = 2$}
  \end{subfigure}
  \begin{subfigure}{0.22\textwidth}
    \centering
    \includegraphics[width=\linewidth]{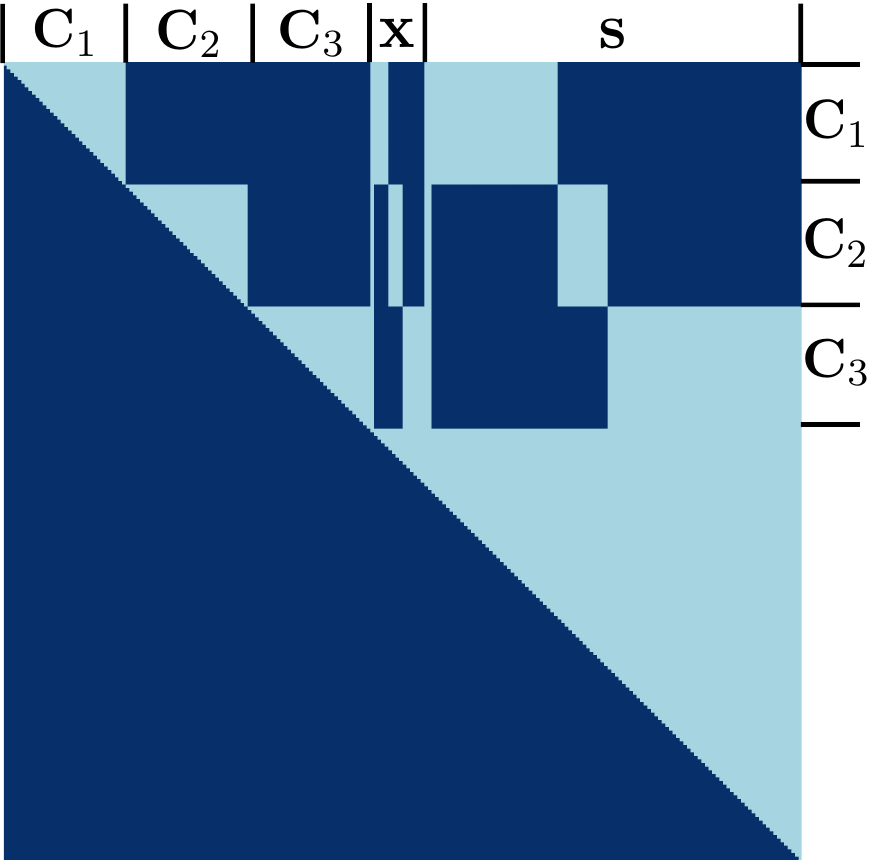}
    \caption{$M = 3$}
  \end{subfigure}
  \begin{subfigure}{0.22\textwidth}
    \centering
    \includegraphics[width=\linewidth]{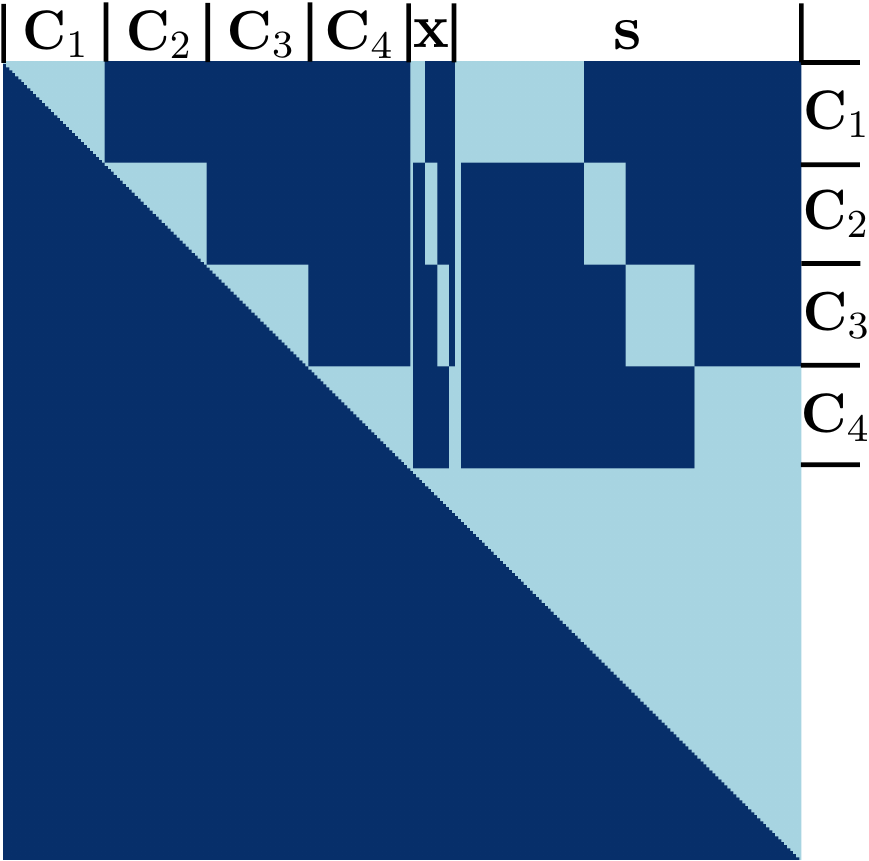}
    \caption{$M = 4$}
  \end{subfigure}
  \begin{subfigure}{0.22\textwidth}
    \centering
    \includegraphics[width=\linewidth]{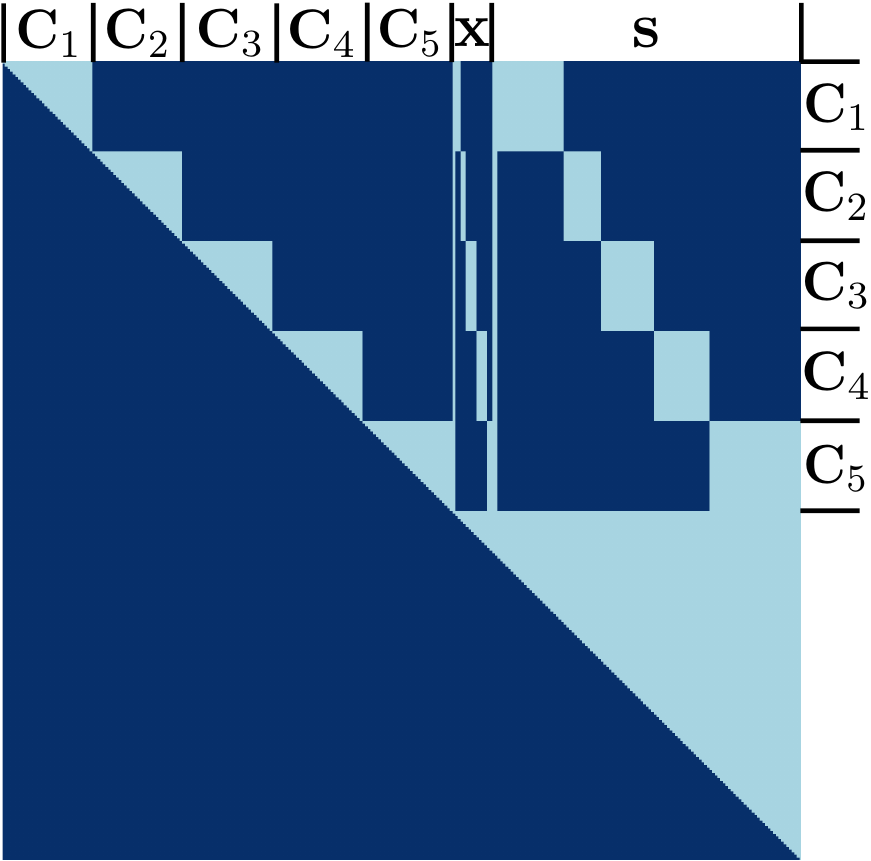}
    \caption{$M = 5$}
  \end{subfigure}
  \caption{Visualization of the final attention mask under varying numbers of segment conditions ($M$).}
  \label{fig:msa_mask_vis}
\end{figure*}

% =========================
% Explanation for Alg.~\ref{alg:emotion_control} (updated for the new pseudocode)
% =========================

This section provides a brief explanation of the symbols and steps used in Alg.~\ref{alg:emotion_control}.

\paragraph{Inputs and outputs.}
$\mathbf{x}=\{x_1, x_2, \ldots,x_T\}$ denotes the source text token sequence of length $T$.
$\mathbf{b}=\{b_1, b_2, \ldots,b_M\}$ are segment boundaries on the text timeline, where $M$ is the number of segments.
$\mathbf{C} = \{\mathbf{C}_1, \mathbf{C}_2, \ldots,\mathbf{C}_M\}$ are the segment-wise condition embeddings (e.g., emotion/style prompts), and each condition may correspond to a short token span of length $L_C$ in the decoder input.
The algorithm autoregressively generates a semantic token stream $\mathbf{s}=\{s_1,s_2,\ldots s_N\}$.

\paragraph{Segment index arrays.}
$\mathrm{seg}_{\mathbf{x}}[t]$ is the segment id assigned to the $t$-th text token $x_t$ according to the boundaries.
$\mathrm{seg}_{\mathbf{s}}[i]$ stores the segment id used when generating the $i$-th semantic token $s_i$. The scalar $m$ denotes the index of the \emph{currently active} segment during decoding and determines which condition embedding is visible to the current generation step.

\paragraph{Step 1: Direct construction of the 2D additive-bias mask $\mathcal{M}_i$.}
At decoding step $i$, the decoder input is organized as a single concatenated token list:
first all segment conditions, then the full text tokens, and finally the already-generated semantic tokens.
Accordingly, the total query/key length is $q = M\!\times\!L_C + T + i$.
We directly build an additive-bias mask $\mathcal{M}_i \in \mathbb{R}^{q\times q}$, where each entry is either $0$ (visible) or $-\infty$ (masked).
Compared to the previous block-matrix presentation, this version writes all constraints as in-place updates on $\mathcal{M}_i$ with explicit offsets for the condition/text/semantic regions.

\begin{itemize}
    \item \textbf{Standard causal visibility.}
    We first initialize $\mathcal{M}_i$ and apply a standard causal mask so that each query token can only attend to itself and earlier tokens in the concatenated sequence. This ensures autoregressive consistency for semantic generation.
    
    \item \textbf{Text-to-condition visibility (segment-local control).}
    For each text token $x_t$, we overwrite its attention row to the \emph{condition region} so that $x_t$ can only see the condition tokens belonging to its own segment.
    Concretely, the entire condition region is masked out for that row, then only the span corresponding to $\mathbf{C}_{\mathrm{seg}_{\mathbf{x}}[t]}$ is unmasked.
    This prevents text tokens in one segment from reading condition prompts from other segments.

    \item \textbf{Semantic-to-condition visibility (segment-local control).}
    Similarly, for each previously generated semantic token $s_r$, we restrict its visibility to the condition region to be segment-local.
    The semantic token can only attend to the condition tokens of the segment recorded in $\mathrm{seg}_{\mathbf{s}}[r]$.
    This enforces that past semantic tokens do not leak information from conditions of unrelated segments.

    \item \textbf{Condition-to-condition isolation (no cross-condition leakage).}
    Condition tokens are not allowed to exchange information across different segments.
    We therefore mask each condition block's attention to all other condition blocks, keeping only the within-block (diagonal) visibility.
    This makes each segment condition self-contained while still allowing the overall model to read text/semantic context under the global causal structure.
\end{itemize}

\paragraph{Step 2: One-step decoding and attention observation.}
Given $\mathcal{M}_i$, the decoder performs one autoregressive step to produce the next semantic token $\mathbf{s}_i$.
During the same forward pass, it also returns the raw attention maps $\mathbf{A}_i$ used as an online alignment observation.
After generation, we append $s_i$ to $\mathbf{s}$ and record $\mathrm{seg}_{\mathbf{s}}[i]\leftarrow m$.

\paragraph{Step 3: Monotonic Stream Alignment (MSA).}
MSA tracks where the semantic stream is aligned on the text in an online manner.
It maintains a posterior belief over text positions and advances it with a monotonic prior transition to encourage forward progression.
From the returned attentions $\mathbf{A}_i$, the algorithm selects a single layer/head whose attention pattern best matches the prior, optionally smooths it to reduce noise, and fuses it with the prior to obtain a stable posterior belief for the current step. In our implementation, the transition factor in $\mathcal{P}$ is set to $p = 0.1$, and a Gaussian smoothing function $\mathcal{G}_{\sigma}$ with $\sigma = 1.2$ is applied.

\paragraph{Step 4: Segment switching.}
The active segment index $m$ is updated by monitoring the expected aligned text position under the current posterior belief.
Once this expected position passes the boundary of the current segment, we increment $m$ to trigger an emotion/style switch for subsequent semantic tokens.
Overall, this mechanism enforces segment-local control through restricted condition visibility, while preserving global coherence via standard causal decoding and online monotonic alignment.

\paragraph{Mask visualization.}
Fig.~\ref{fig:msa_mask_vis} illustrates the resulting mask pattern produced by the direct in-place construction in Step~1 when the utterance is partitioned into different numbers of segments ($M{=}2,3,4, 5$).
Each panel shows how the condition token blocks, text tokens, and semantic tokens are jointly constrained by (i) the global causal structure and (ii) the segment-local condition visibility.
As $M$ increases, the condition region is divided into more isolated blocks, and each text/semantic token is restricted to attend only to the condition block of its assigned (or currently active) segment.
This visualization helps verify that the mask enforces local emotion/style control without allowing cross-condition leakage across segments. 

\section{Segment-Aware Duration Steering}
\label{appendix:C}
In practice, segment-aware duration steering is implemented as two lightweight inference-time controllers that operate entirely on semantic token counts and alignment signals. For local duration steering, a proportional controller described in Section~\ref{subsec:multi_segment_duration_control} performs online correction by comparing the normalized semantic generation progress within the active segment to the normalized text progress obtained from the MSA algorithm. The correction is applied with gain $k_p = 25.0$, which determines the sensitivity of duration adjustment to progress mismatch, and is triggered only when the absolute progress error exceeds $\varepsilon = 0.01$, thereby preventing unnecessary updates caused by minor alignment fluctuations. To ensure stability, updates are performed at a fixed low frequency of one update every five decoding steps, and the per-update adjustment is clamped to a maximum magnitude of $\Delta_{\max} = 10$ semantic tokens to avoid abrupt changes in generation pace. In addition, the effective target is constrained by adaptive lower bounds tied to the current global decoding cursor, with conservative and emergency regimes activated when the generated length exceeds $1.2\times$ and $1.5\times$ the planned segment budget, respectively, serving as a safeguard against uncontrolled over-generation. 

For global EOS steering, an EOS controller is added to the logits processor list, where EOS logits are fully suppressed for all non-final segments, while being dynamically adjusted in the final segment based on the ratio between generated semantic tokens and the target semantic budget. Specifically, EOS is strongly suppressed when the ratio is below $0.5$, gradually transitions to a neutral region over the interval $[0.8, 1.1]$, and is increasingly encouraged as the ratio approaches $1.2$, with the applied bias bounded between $-5.0$ and $+15.0$. These fixed hyperparameters were selected empirically and remain constant across all experiments, enabling robust intra-utterance duration control without modifying or retraining the underlying TTS model.

\section{Ablation Models Implementation}
\label{appendix:D}
\subsection{Emotion and Duration Control}
In the segment-aware emotion conditioning part of Tab.~\ref{abl_1}, we compare our method with the following two ablated variants:
\begin{itemize}
  \item \textbf{w/o full-text access}: In this variant, each segment condition can only attend to the local text tokens within its own segment, rather than the full text.
  \item \textbf{w/o alignment}: In this variant, we remove any alignment module and generate semantic tokens by randomly switching phases through a fixed probability at each step.
\end{itemize}

For the segment-aware duration steering part of Tab.~\ref{abl_1}, we further evaluate two ablated variants to analyze the contributions of local and global steering mechanisms:
\begin{itemize}
  \item \textbf{w/o local steering}: In this variant, the local duration steering module is disabled, and segment-level pacing relies solely on the baseline duration embedding, while the global EOS control mechanism is retained.
  \item \textbf{w/o global EOS}: In this variant, the global EOS logit modulation is disabled, while the local duration steering module remains active.
\end{itemize}

\subsection{Monotonic Stream Alignment Evaluation}
\label{appendix:D_MSA_abl}
In Fig.~\ref{abl_3}, we compare our MSA method with the following ablated variants:
\begin{itemize}
  \item \textbf{Max Attention Head + Greedy Monotonic Alignment}: In this variant, 
  we replace our MSA with a deterministic heuristic. We firstly compute a score for each raw attention maps across all layers and heads through $F^{(l,h)} = \frac{1}{T} \sum_{t=1}^{T} \mathbf{A}^{(l,h)}_{i,t}$, where $T$ is the length of text tokens and $t$ is the text position. The optimal attention map $(l^*, h^*)$ is selected as the observation by the maximum score. For the update step, we restrict a monotonic constraint and simplify the posterior $\boldsymbol{\pi}_i$ to a one-hot vector, representing a hard alignment state. Let $k$ be the active index at the previous step, i.e., $\boldsymbol{\pi}_{i-1} (k) = 1$, and the update rule follows a greedy local comparison between the current position $k$ and the next position $k+1$. The new belief is determined as:
  \begin{equation}
    \boldsymbol{\pi}_i(t) = \mathds{1}\left[ t = \operatorname*{arg\,max}_{m\in \{k, k+1\}} \mathbf{A}^{(l, h)}_{i,m} \right], 
  \end{equation}
  where $\mathds{1}[\cdot]$ is the indicator function.
  \item \textbf{Top-$k$ Attention Heads + Greedy Monotonic Alignment}: In this variant, we extend the previous method by selecting the top-$k$ attention heads as observations. Specifically, we first compute the scores $F^{(l,h)}$ for all attention maps and select the top-$k$ heads with the highest scores. The observation is then derived as a weighted average of these selected attention maps based on their scores. The greedy monotonic alignment update remains the same as above.

  \item \textbf{Max Attention Head + Monotonic Stream Alignment}: In this variant, we retain alignment updates using our MSA algorithm. We replace the observation component by selecting a single attention head with the maximum score as described above, and get rid of the smoothing operation.
\end{itemize}

\section{Evaluation Protocol}
\label{appendix:E}
\subsection{Baseline and Comparative models}
\textbf{Baseline.} IndexTTS2\footnote{\url{https://github.com/index-tts/index-tts}}~\cite{zhou2025indextts2} is an autoregressive zero-shot TTS model that supports utterance-level control of emotion and speech duration while maintaining high speech naturalness. It disentangles speaker identity from emotional expression, enabling faithful reconstruction of target timbre and accurate reproduction of the specified emotional style. By incorporating GPT-based latent representations, the model further improves semantic consistency and stability under expressive conditions.

We also adopt several strong zero-shot TTS as our comparative methods:
\begin{itemize}
  \item \textbf{MaskGCT}\footnote{\url{https://github.com/open-mmlab/Amphion/tree/main/models/tts/maskgct}}~\cite{wang2025maskgct} is a non-autoregressive TTS model that a masked generative transformer to predict semantic and acoustic tokens, functioned with duration control. By leveraging two-stage mask prediction mechanism, it achieves high fidelity and robust voice synthesis.
  \item \textbf{F5TTS}\footnote{\url{https://github.com/SWivid/F5-TTS}}~\cite{chen2025f5} is a non-autoregressive TTS system based on Diffusion Transformer (DiT). It eliminates explicit alignment by padding text to speech length. Trained on 100k hours of data, it employs Sway Sampling to achieve efficient, high-quality zero-shot multilingual synthesis.
  \item \textbf{SparkTTS}\footnote{\url{https://github.com/SparkAudio/Spark-TTS}}~\cite{wang2025spark} is a powerful TTS system built upon Qwen2.5, which directly reconstructs audio from LLM-predicted codes and eliminates the need for complex intermediate models like flow matching. It excels in high-fidelity zero-shot voice cloning for bilingual scenarios while maintaining high efficiency.
  \item \textbf{CosyVoice2}\footnote{\url{https://github.com/FunAudioLLM/CosyVoice?tab=readme-ov-file}}~\cite{du2024cosyvoice2} is an autoregressive TTS model that combines a language model for semantic and prosodic modeling with flow matching for speaker identity reconstruction, utilizing a supervised speech tokenizer to achieve disentangled generation. Notably, it demonstrates superior performance in Chinese compared to English due to its training data distribution.
\end{itemize}

Our baseline and comparative models adopt a consistent segment-wise inference strategy. Each sentence is partitioned into multiple segments based on target emotions and speaking rates generated by our fine-tuned LLM. These segments are generated individually among these models and sequentially assembled to reconstruct the complete utterance for evaluation. All baseline and comparative models are implemented using their official open-source codebases and pretrained weights.

\subsection{Subjective Evaluation}
\begin{figure*}[ht]
  \centering
  \begin{subfigure}[b]{\textwidth}
    \centering
    \includegraphics[width=\textwidth]{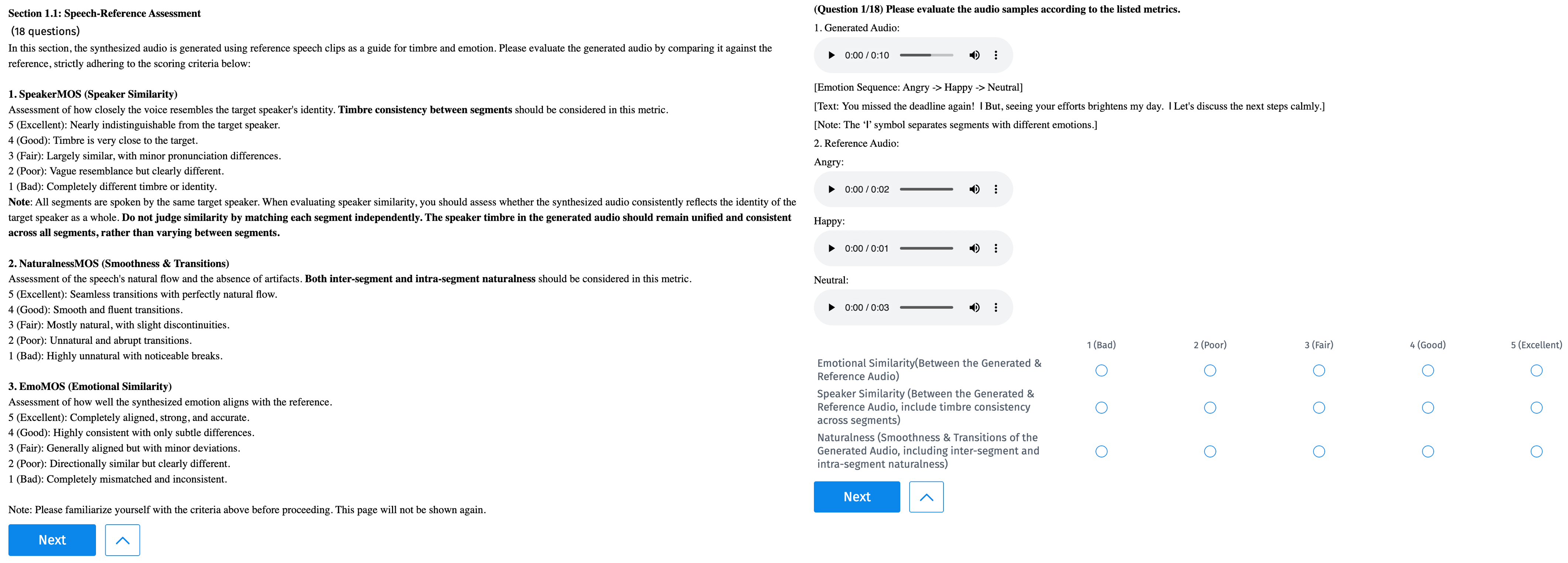}
    \caption{Speech-Prompted Emotion Control Evaluation}
    \label{fig:mos-online-a}
  \end{subfigure}
  \hfill
  \begin{subfigure}[b]{\textwidth}
    \centering
    \includegraphics[width=\textwidth]{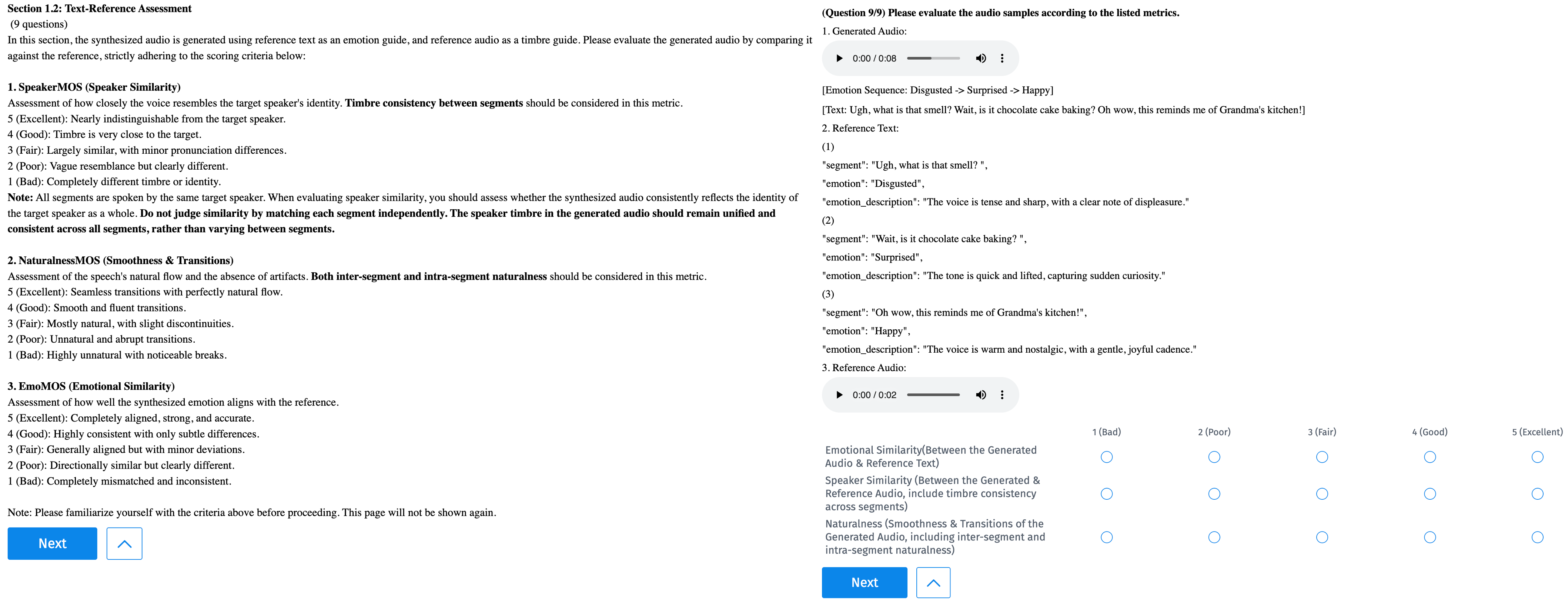}
    \caption{Text-Prompted Emotion Control Evaluation}
    \label{fig:mos-online-b}
  \end{subfigure}
  \begin{subfigure}[b]{\textwidth}
    \centering
    \includegraphics[width=\textwidth]{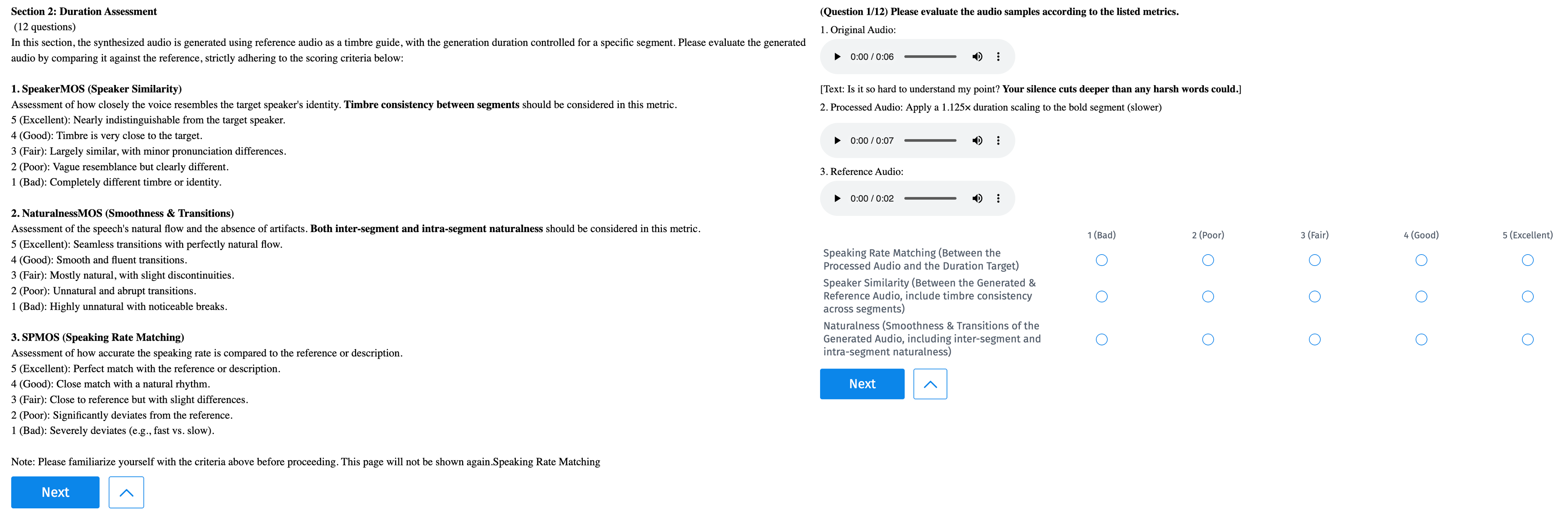}
    \caption{Duration Control Evaluation}
    \label{fig:mos-online-c}
  \end{subfigure}
  \caption{User interface for MOS evaluation across different evaluation tasks.}
  \label{fig:mos-online}
\end{figure*}

We conduct a subjective Mean Opinion Score (MOS) evaluation focusing on four key dimensions: emotion consistency, speaking rate consistency, speaker similarity, and emotional transition smoothness. Participants were provided with explicit scoring criteria, and we report the mean scores along with 95\% confidence intervals (CI) in Tab.\ref{com_1} and Tab.\ref{com_2}. The evaluation involved 15 graduate students with relevant research backgrounds. Prior to the evaluation, participants were provided with detailed task protocols and informed of the specific usage of the data. Each participant evaluated 18 test samples (9 Chinese and 9 English) under different settings, with the entire session lasting approximately 40 minutes. Scores ranged from 1 to 5 with 1-point intervals. Each participant received compensation of 15 SGD for their participation. The user interface for MOS evaluation is illustrated in Fig.~\ref{fig:mos-online}.

\subsection{Objective Evaluation}
Our objective evaluation encompasses several metrics to assess various aspects of speech synthesis quality and controllability. Character accuracy is measured using an automatic speech recognition (ASR) model through comparison with ground-truth transcriptions. For English audio evaluation, we employ a Whisper Large V3~\cite{radford2023robust} ASR model to calculate Word Error Rate (WER)\footnote{\url{https://huggingface.co/openai/whisper-large-v3}}, while for Chinese audio, we utilize a Paraformer~\cite{gao2022paraformer} ASR model to calculate Character Error Rate (CER) for Chinese to quantify transcription accuracy\footnote{\url{https://huggingface.co/funasr/paraformer-zh}}. 

To evaluate the smoothness of transitions in both emotion and speaking rate, we adopt the DNSMOS Pro\footnote{\url{https://github.com/fcumlin/DNSMOSPro}}~\cite{cumlin2024dnsmos}, referred as DNSM. It is calculated by averaging the predicted MOS values obtained from a sliding window (2-second duration, 1-second stride) applied across the full utterance. Speaker similarity is assessed using fine-tuned WavLM-Large~\cite{chen2022wavlm} for speaker verification\footnote{\url{https://github.com/microsoft/UniSpeech/tree/main/downstreams/speaker_verification}} to extract speaker embeddings from synthesized and reference audios, followed by computing the cosine similarity, denoted as SSIM. We report the average scores of the two metrics: the similarity between the synthesized and reference audios, and the intra-utterance consistency measured across all segment pairs obtained via ASR-based segmentation within the generated speech. 

For speech naturalness evaluation, we utilize NISQA\footnote{\url{https://github.com/gabrielmittag/NISQA}}~\cite{mittag2021nisqa} and OVRL from DNSMOS\footnote{\url{https://github.com/microsoft/DNS-Challenge/tree/master/DNSMOS}}~\cite{reddy2022dnsmos} for overall quality of a synthesized sequence. Both of them are evaluated through the entire utterance without segmentation. 
The emotional expression accuracy is measured through extracting segment-level emotional embeddings from ASR-segmented audio clips using a pre-trained speech emotion recognition model emotion2vec-large\footnote{\url{https://huggingface.co/emotion2vec/emotion2vec_plus_large}}~\cite{ma2024emotion2vec}. We calculate the cosine similarity between synthesized and reference audios for in speech prompt settings, and utilize classification accuracy over 5 discrete emotional labels for text prompt settings.

\subsection{Experimental Result Supplements}
% ----------------------- Speech Input English Category Table -----------------------
\begin{table*}[t!]
\centering
\resizebox{0.95\textwidth}{!}{%
\begin{threeparttable}
\begin{tabular}{c l l c c c c c c}
\toprule
\textbf{Prompt} & 
\textbf{Category} & 
\textbf{Method} & 
\textbf{WER$\downarrow$} & 
\textbf{DNSM$\uparrow$} & 
\textbf{SSIM$\uparrow$} & 
\textbf{NISQA$\uparrow$} & 
\textbf{OVRL$\uparrow$} & 
\textbf{Emo2vec$\uparrow$} \\
\midrule

% ====================================================================
% =========================== SPEECH INPUT ===========================
% ====================================================================
\multirow{9}{*}{Speech} & 
% --- Emotional Dialogue ---
\multirow{3}{*}{Emotional Dialogue}
  & MaskGCT   & 1.978 & 3.757 & 0.331 & 4.375 & 3.266 & \textbf{0.862} \\
  & & CosyVoice & \textbf{0.459} & 3.570 & 0.394 & 4.403 & 3.321 & 0.822 \\
  \rowcolor{gray!15} \cellcolor{white} & \cellcolor{white} & Ours      & 5.462 & \textbf{3.839} & \textbf{0.442} & \textbf{4.670} & \textbf{3.372} & 0.811 \\
\cmidrule{2-9}

% --- Observational Phrase ---
 & \multirow{3}{*}{Observational Phrase}
  & MaskGCT   & 4.468 & 3.849 & 0.352 & 4.518 & 3.277 & \textbf{0.854} \\
  & & CosyVoice & 1.021 & 3.567 & 0.410 & 4.515 & 3.333 & 0.844 \\
  \rowcolor{gray!15} \cellcolor{white} & \cellcolor{white} & Ours      & \textbf{0.834} & \textbf{3.903} & \textbf{0.491} & \textbf{4.710} & \textbf{3.418} & 0.848 \\
\cmidrule{2-9}

% --- Vivid Descriptive ---
 & \multirow{3}{*}{Vivid Description}
  & MaskGCT   & 3.612 & 3.866 & 0.377 & 4.463 & 3.267 & 0.831 \\
  & & CosyVoice & \textbf{0.834} & 3.621 & 0.443 & 4.535 & 3.313 & 0.826 \\
  \rowcolor{gray!15} \cellcolor{white} & \cellcolor{white} & Ours      & 1.085 & \textbf{3.940} & \textbf{0.508} & \textbf{4.636} & \textbf{3.391} & \textbf{0.836} \\

\midrule

% ====================================================================
% ============================ TEXT INPUT ============================
% ====================================================================
\multirow{9}{*}{Text} & 
% --- Emotional Dialogue ---
\multirow{3}{*}{Emotional Dialogue}
  & CosyVoice     & \textbf{0.956} & 3.450 & \textbf{0.473} & 4.336 & 3.294 & 0.369 \\
  & & IndexTTS2  & 4.059 & 3.506 & 0.399 & 4.201 & 3.194 & \textbf{0.596} \\
  \rowcolor{gray!15} \cellcolor{white} & \cellcolor{white} & Ours          & 5.996 & \textbf{3.634} & 0.387 & \textbf{4.515} & \textbf{3.324} & 0.468 \\
\cmidrule{2-9}

% --- Observational Phrase ---
 & \multirow{3}{*}{Observational Phrase}
  & CosyVoice     & 2.582 & 3.464 & 0.435 & 4.385 & 3.294 & 0.270 \\
  & & IndexTTS2  & \textbf{0.747} & 3.531 & 0.436 & 4.316 & 3.220 & \textbf{0.416} \\
  \rowcolor{gray!15} \cellcolor{white} & \cellcolor{white} & Ours & 1.104 & \textbf{3.657} & \textbf{0.477} & \textbf{4.606} & \textbf{3.327} & 0.381 \\
\cmidrule{2-9}

% --- Vivid Descriptive ---
 & \multirow{3}{*}{Vivid Description}
  & CosyVoice     & 1.631 & 3.483 & 0.437 & 4.373 & 3.284 & 0.289 \\
  & & IndexTTS2  & 1.442 & 3.511 & 0.450 & 4.353 & 3.235 & \textbf{0.489} \\
  \rowcolor{gray!15} \cellcolor{white} & \cellcolor{white} & Ours          & \textbf{0.756} & \textbf{3.762} & \textbf{0.486} & \textbf{4.644} & \textbf{3.372} & 0.393 \\

\bottomrule
\end{tabular}
\end{threeparttable}
}
\caption{Objective evaluation results on English Speech and Text inputs across different text categories. $\downarrow$ indicates that lower values are better, while $\uparrow$ indicates that higher values are better. Best results are \textbf{bolded}.}
\label{tab:consolidated_results}
\end{table*}

\begin{figure}[t!]
  \centering
  \includegraphics[width=\columnwidth]{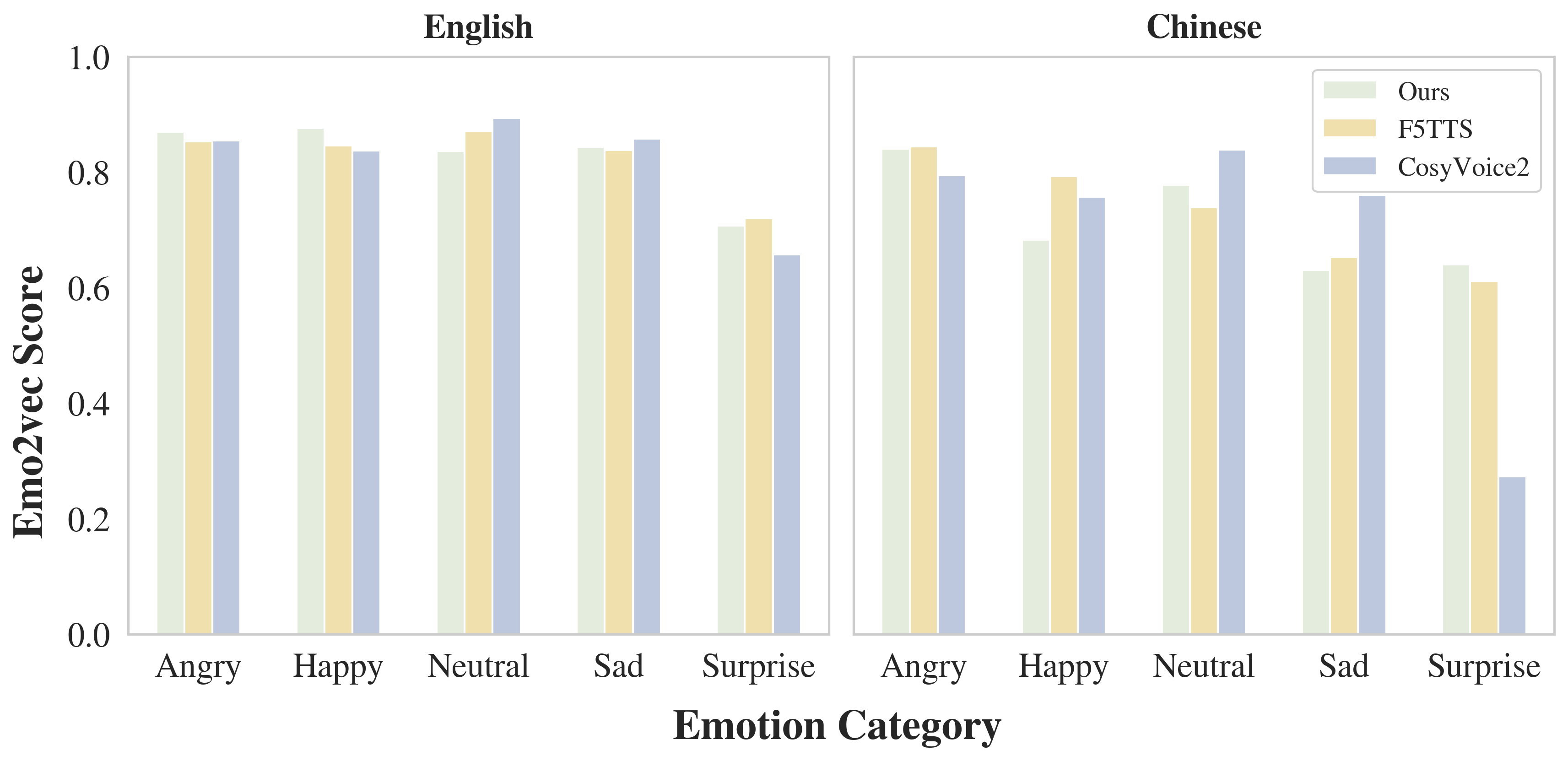}
  \caption{Comparison of Emo2Vec similarity scores across languages for five emotion categories. Our method is compared with F5TTS\cite{chen2025f5} and CosyVoice2\cite{du2024cosyvoice2}.}
  \label{fig:emo2vec_cat}
\end{figure}

% ----------------------- Alignment Strategy Comparison Table -----------------------
\begin{table}[t]
\centering
\footnotesize
\resizebox{\linewidth}{!}{
\begin{tabular}{@{}>{\hspace{1mm}}lccccc@{}}
\toprule
\textbf{Method} &
\textbf{DNSM$\uparrow$} &
\textbf{SSIM$\uparrow$} &
\textbf{NISQA$\uparrow$} &
\textbf{OVRL$\uparrow$} &
\textbf{Emo2vec$\uparrow$} \\
\midrule
\rowcolor{gray!15} \textbf{Ours} & \textbf{3.925} & \textbf{0.485} & \textbf{4.706} & \textbf{3.395} & \textbf{0.837} \\
Max Head + Greedy           & 3.901 & 0.443 & 4.683 & 3.372 & 0.803 \\
Top-$k$ + Greedy     & 3.878 & 0.442 & 4.702 & 3.383 & 0.815 \\
Max Head + MSA                & 3.907 & 0.462 & 4.697 & 3.393 & 0.828 \\

\bottomrule
\end{tabular}
}
\vspace{-5pt}
\caption{Objective Comparison of Different Alignment Strategies. $\uparrow$ indicates that higher values are better. Best results are \textbf{bolded}.}
\label{tab:alignment_comparison}
\end{table}

\noindent \textbf{Emotion-Specific Control Evaluation.}
We provide detailed experimental results for the emotional similarity across five discrete categories (Angry, Happy, Neutral, Sad, Surprise). Our results are shown in Fig.~\ref{fig:emo2vec_cat}, highlighting distinct performance patterns across different input modalities. Our method demonstrates robust emotional fidelity, achieving emotional similarity scores that closely approach those of comparative methods. It's worth emphasizing that while other methods could only generate single emotion clips where the global style is constant, our method generates continuous and multi-segment sequences with transitioning emotions. Despite the difficulty of modeling such dynamic emotional control, our model still maintains high emotional similarity, and even superior performance in certain categories such as Angry and Sad, proving its effectiveness in generating complex and varying prosody. 

\noindent \textbf{Category-Specific Emotion Control Evaluation.}
Tab.~\ref{tab:consolidated_results} extends our evaluation to three distinct synthesis scenarios in our dataset: Emotional Dialogue, Observational Phrase, and Vivid Description. Our method demonstrates a consistent advantage in audio quality, achieving the highest NISQA and OVRL scores in almost all settings. It also excels in naturalness of emotional transitions, as reflected by the DNSM metric, where our method consistently outperforms all methods across different text categories and input modalities.
This further confirms that our segment-aware generation effectively maintains naturalistic acoustic synthesis even when handling complex emotional transitions.

Nevertheless, we observe that our method still faces challenges in speaker similarity in certain scenarios. In the text prompt setting, our method lags behind CosyVoice2 in Emotional Dialogue, where the generated sentences often contain extensive emotionally charged and oral conversational elements, such as modal particles and emphatic punctuation. In this scenario, our method and baseline prioritize the expressive prosody, which may lead to deviations from the target speaker's timbre. While CosyVoice2 keeps a more stable and consistent prosody during generation, and preserves speaker identity, it fails to convey the intended emotional expressions.
This highlights the inherent trade-off between emotional expressiveness and speaker fidelity in zero-shot TTS, especially when generating highly dynamic prosody from text alone. 

\noindent \textbf{MSA Ablation Studies.}
We further validate the effectiveness of our proposed Monotonic Stream Alignment (MSA) in Tab.~\ref{tab:alignment_comparison}. We compare our MSA-based alignment strategy against several variants illustrated in Appendix~\ref{appendix:D_MSA_abl}. These protocols are the same as those in Fig.~\ref{abl_3}. The results show that replacing our MSA with greedy monotonic alignment leads to noticeable performance drops across all metrics, indicating that the MSA mechanism is crucial not just for text-audio synchronization, but also for stabilizing the emotional contents. Through maintaining a robust posterior belief of the current position, MAS prevents the model from drifting off the complex segment boundaries, thereby ensuring the naturalness and coherence of emotional transitions. Notably, even without the full MSA mechanism, these ablation variants still maintain relatively high performance levels. This suggests that synthesizing speech containing multiple complex emotions in a single continuous streaming process, rather than generating each segment independently, inherently preserves semantic and acoustic coherence, which benefits the overall quality of the generated speech.

\begin{figure*}[t]
\centering
\begin{minipage}{\textwidth}

\begin{lstlisting}[
  basicstyle=\ttfamily\footnotesize,
  frame=single,
  breaklines=true,
  breakatwhitespace=true,
  columns=fullflexible,
  keepspaces=true,
  showstringspaces=false,
  linewidth=\textwidth,
  xleftmargin=0.5em,
  xrightmargin=0.5em
]

Role:
You are an expert creative screenwriter and emotional expression specialist.
Your task is to generate high-quality text utterances for text-to-speech synthesis evaluation.

Task: 
Given an ordered emotion sequence, generate a single-sentence text utterance that reflects 
the emotional journey described by the sequence. 
The text should naturally transition through these emotions in order.

Emotion Sequence:
1. ${Emotion_1}$
2. ${Emotion_2}$
3. ${Emotion_3}$

Requirements:

1. Text Utterance:
   - Length: 15-25 words (corresponding to 5-10 seconds of speech).
   - The text MUST contain all emotions in the given sequence, each clearly identifiable.
   - Emotional transitions MUST be conveyed through changes in language tone, imagery, internal reactions, or perspective.
   - CRITICAL: Do NOT use explicit temporal markers such as "then", "now", "afterward", "at first", "later", "next", "suddenly", or "finally".
   - The sentence must be semantically coherent and flow naturally as a single utterance.
   - Avoid clich\'ed or overused expressions, especially as opening phrases.
   - The opening MUST be unique and creative; avoid common narrative patterns.

2. Text Category Constraint:
   ${
   - vivid_descriptive: Vivid descriptive sentences (novel prose style). Example: "Wind whispered through the parched cornstalks, its voice fraying like worn silk." |
   - emotional_dialogue: Emotionally charged dialogue excerpts (natural spoken lines). Example: "I've asked you three times! Why is the door still locked?" |
   - observational_phrase: Observational phrases (subtle situational commentary). Example: "Rain taps the window like it's bruising the glass-rhythmic, insistent, all night."
   }$

3. Output Format:
   Provide your response in the following JSON structure ONLY:
   {
     "text": "<generated single-sentence utterance>",
     "text_category": "${text_category: vivid_descriptive | emotional_dialogue | observational_phrase}$"
   }

Examples:

Example 1 
${Example: 
Vivid Descriptive 
Input Emotion Sequence:
1. Happy
2. Surprised
3. Sad
Output:
{
  "text": "Warm light drifts around me, a sudden sharp gust jolts the calm,
           and a muted heaviness settles quietly over my thoughts.",
  "text_category": "vivid_descriptive"
}
}$
...

Now generate a text utterance for the given emotion sequence.
\end{lstlisting}

\end{minipage}
\captionsetup{type=lstlisting}
\caption{Example prompt for generating content text with emotion shifts using GPT-4o.}
\label{lst1}
\end{figure*}

\begin{figure*}[t]
\centering
\begin{minipage}{\textwidth}

\begin{lstlisting}[
  basicstyle=\ttfamily\footnotesize,
  frame=single,
  breaklines=true,
  breakatwhitespace=true,
  columns=fullflexible,
  keepspaces=true,
  showstringspaces=false,
  linewidth=\textwidth,
  xleftmargin=0.5em,
  xrightmargin=0.5em
]
Role:
You are an expert linguistic annotator specialized in emotional prosody for TTS datasets.
Your task is to segment the given sentence into emotion-aligned segments while preserving the exact
original wording.

Task:
Segment the following text into contiguous spans that correspond to the emotions in the sequence. 
Each segment must represent a natural linguistic unit and reflect its assigned emotion 
through tone, sensory cues, or attitude-NOT through explicit time markers.

Input Text:
${Original text generated in Step-1}$

Emotion Sequence:
1. ${Emotion_1}$
2. ${Emotion_2}$
3. ${Emotion_3}$

Requirements:

1. Segmentation Rules:
   - Produce EXACTLY the same number of segments as emotions in the sequence.
   - CRITICAL: Segments MUST correspond to the emotion sequence IN ORDER.
     The first segment maps to the first emotion, the second to the second emotion, etc.
   - Each segment MUST be a continuous span from the original text.
     Do NOT rewrite, reorder, omit, or add any words.
   - All punctuation marks from the original text MUST be preserved in their exact positions.
   - The concatenation of all segments MUST reconstruct the original text exactly.
   - Segment boundaries should align with natural linguistic or prosodic boundaries
     (e.g., phrase or clause boundaries). Do NOT split inside tight phrases.

2. Emotion Description (for TTS prosody reference):
   - Provide a short vocal-affect description (5-15 words) focusing on auditory qualities.
   - The description should focus on auditory characteristics (e.g., pitch,
     intensity, pacing), not on events or semantics.
   - The description MUST align with the assigned emotion.

3. Speaking Time Estimation:
   - Estimate speaking duration in seconds using the guideline:
     0.18-0.30 seconds per word as a baseline.
   - The estimated duration should also reflect the emotional tone of the segment, as different emotions naturally influence speaking pace (e.g., excited or tense delivery tends to be quicker, while somber or reflective delivery tends to slow down).
   - The final time MUST be a realistic approximation of how the segment would be delivered aloud.
   - IMPORTANT: The sum of all segment durations MUST fall within 5-13 seconds.
   - Output time values as decimal strings (e.g., "2.4").

Output Format (JSON ONLY):
{
  "original_text": "${original input text}$",
  "segments": [
    {
      "lines_seg": "<text segment>",
      "emotion": "<emotion label from the sequence>",
      "emotion_description": "<vocal-affect description>",
      "time": "<estimated speaking time in seconds>"
    },
    ...
  ]
}

Example:
...

Now generate the segmentation for the given input text and emotion sequence.
\end{lstlisting}

\end{minipage}
\captionsetup{type=lstlisting}
\caption{Example prompt for emotion-aligned segmentation and duration annotation using DeepSeek-Chat.}
\label{lst2}
\end{figure*}

\begin{figure*}[t]
\centering
\begin{minipage}{\textwidth}

\begin{lstlisting}[
  basicstyle=\ttfamily\footnotesize,
  frame=single,
  breaklines=true,
  breakatwhitespace=true,
  columns=fullflexible,
  keepspaces=true,
  showstringspaces=false,
  linewidth=\textwidth,
  xleftmargin=0.5em,
  xrightmargin=0.5em
]
Manual Review Checklist (total 1,000 samples: 500 EN / 500 ZH)

----------------------------------------
[Step 1] Content Text Generation

- Text validity:
  the text is complete, fluent, and natural, without obvious truncation, repetition, or unfinished clauses (typically a single well-formed sentence).
- Length appropriateness:
  text length falls within the intended range (EN: 15-25 words; ZH: 15-30 characters), and does not appear unnaturally compressed or padded to meet length requirements.
- Semantic coherence:
  the text conveys a single coherent idea or situation, rather than a loose collection of phrases or unrelated clauses.
- Category consistency:
  the assigned text category matches the content style (vivid descriptive / emotional dialogue / observational phrase), with category cues clearly identifiable within the text.
- Emotion sequence correctness:
  the emotion sequence contains 2-3 valid emotions drawn from the predefined set, and all emotions are meaningfully reflected somewhere in the text.
- Emotion progression naturalness:
  emotional transitions implied by the text occur in a plausible order, without abrupt or logically unsupported emotion jumps.
- Language quality:
  the text does not contain obvious grammatical errors, unnatural phrasing, or machine-like constructions that would hinder natural speech rendering.

----------------------------------------
[Step 2] Multi-segment Prompt Annotation

- Segmentation boundaries:
  segment splits occur at natural linguistic or prosodic boundaries, such as phrase or clause breaks, and avoid splitting fixed expressions or tight collocations.
- Emotion-text alignment:
  the semantic content of each segment clearly supports its assigned emotion, and the intended emotion is perceivable without relying on the description.
- Vocal-affect specificity:
  emotion description includes concrete auditory cues (e.g., energy level, pitch tendency, speaking rate, intensity) rather than abstract emotion names.
- Description naturalness:
  emotion description reads as a natural speaking instruction and typically spans one short phrase or sentence, rather than a list of keywords.
- Duration plausibility:
  estimated speaking durations are reasonable given segment length and linguistic complexity, and fall within the expected range of 0.3-8.0 seconds per segment.
- Duration consistency:
  duration differences across segments reflect intuitive pacing differences, such as faster delivery for excited emotions and slower delivery for calm or reflective ones.
- Coverage consistency:
  concatenated segment texts fully reconstruct the original text, with no missing, duplicated, or reordered content.

----------------------------------------
Manual Review Protocol

- Reviewers:
  all sampled items are independently inspected by at least two reviewers, covering both English and Chinese samples.
- Disagreement handling:
  cases with inconsistent judgments are discussed and resolved through consensus review, and recurring issues are recorded for prompt or rule refinement.


\end{lstlisting}

\end{minipage}
\captionsetup{type=lstlisting}
\caption{Manual verification checklist used in our human review process for Step~1 and Step~2 outputs.}
\label{lst3}
\end{figure*}

\begin{algorithm*}[t]
\small
\SetAlgoLined
\DontPrintSemicolon
\SetKwInOut{Input}{Input}
\SetKwInOut{Output}{Output}
\SetKwFunction{G}{\mathcal{G}_{\sigma}}

\caption{Segment-Aware Emotion Conditioning with Monotonic Stream Alignment (MSA)}
\label{alg:emotion_control}

\Input{Text tokens $\mathbf{x}=\{x_1, x_2, \ldots,x_T\}$, segment boundaries $\mathbf{b}=\{b_1, b_2, \ldots,b_M\}$, condition embeddings $\mathbf{C} = \{\mathbf{C}_1, \mathbf{C}_2, \ldots,\mathbf{C}_M\}$}
\Output{Generated semantic tokens $\mathbf{s}=\{s_1,s_2,\dots,s_N\}$}

\tcp{Compute segment id for text tokens $\mathrm{seg}_{\mathbf{x}}[1..T]$:}
\For{$t\leftarrow 1$ \KwTo $T$}{
    $\mathrm{seg}_{\mathbf{x}}[t] \leftarrow 1 + \sum_{r=1}^{M-1} \mathbb{I}[t > b_r]$\;
}

Initialize $\mathrm{seg}_{\mathbf{s}}\leftarrow [\ ]$ \tcp*[r]{store segment id for each generated semantic token}
Initialize semantic index $i \leftarrow 0$, segment index $m \leftarrow 1$, $\mathbf{s}\leftarrow \emptyset$\;
Initialize posterior alignment belief $\boldsymbol{\pi}_0 \in \mathbb{R}^{T}$ (one-hot at $t=1$)\;

\While{not EndOfSentence}{
    $i \leftarrow i + 1$\;

    \tcp{1) Build 2D additive-bias offset mask directly on $\mathcal{M}_i$}

    $q \leftarrow M \times L_C + T + i$ \tcp*[r]{\#Q tokens: $[\mathbf{C}_{1:M}, x_{1:T}, s_{1:i}]$}

    % Keys = [Q, C], Queries = [Q, C]
    $\mathcal{M}_i \leftarrow (-\infty)\cdot \mathbf{1}_{(q)\times(q)}$
    
    \tcp{1.1) Standard causal mask}
    % (a) Q -> Q: standard causal
    \For{$u\leftarrow 1$ \KwTo $q$}{  
      \For{$v\leftarrow 1$ \KwTo $u$}{
        $\mathcal{M}_i[u,v]\leftarrow 0$\;
      }
    }

    \tcp{1.2) x -> C: Text tokens to condition embeddings (segment-local condition visibility)}
    $\,\mathrm{off} \leftarrow M \times L_C\,$
    
    \For{$t\leftarrow 1$ \KwTo $T$}{
    $\mathcal{M}_i[\mathrm{off} +t,\ 0 : M \times L_C] \leftarrow -\infty$\;
    
      $\mathcal{M}_i[\mathrm{off} +t,\ L_C \times (\mathrm{seg}_{\mathbf{x}}[t]-1) : L_C \times \mathrm{seg}_{\mathbf{x}}[t]] \leftarrow 0$\;
    }

    \tcp{1.3) S -> C: Semantic tokens to condition embeddings (segment-local condition visibility)}
    $\,\mathrm{off} \leftarrow M \times L_C + T\,$
    
    \For{$r\leftarrow 1$ \KwTo $i-1$}{
    $\mathcal{M}_i[\mathrm{off} +r,\ 0 : M \times L_C] \leftarrow -\infty$\;
    
      $\mathcal{M}_i[\mathrm{off}+r,\ L_C \times (\mathrm{seg}_{\mathbf{s}}[r]-1) : L_C \times \mathrm{seg}_{\mathbf{s}}[r]] \leftarrow 0$\;
    }

    \tcp{1.4) C -> C: Condition embeddings to condition embeddings (no cross-condition leakage)}
    
    \For{$u\leftarrow 1$ \KwTo $M$}{

        $\mathcal{M}_i[(u-1)  \times L_C : u  \times L_C,\ 0 : M \times L_C] \leftarrow -\infty$\;

        $\mathcal{M}_i[(u-1)  \times L_C : u  \times L_C,\ (u-1)  \times L_C : u  \times L_C] \leftarrow 0$\;
      
    }

    \tcp{2) Decode one step with mask and get raw attentions as observation}
    $(s_i,\ \mathbf{A}_i) \leftarrow
    f_{\theta}^{\text{decode-step}}\!\Big(
    \mathbf{x},\ \mathbf{s}_{<i},\ \{\mathbf{C}_j\}_{j=1}^{M}, \
    \mathcal{M}_i,\ \texttt{return\_attn}=True
    \Big)$\;

    $\mathbf{s}\leftarrow \{\mathbf{s} , \ s_i\}$\;
    
    $\mathrm{seg}_{\mathbf{s}}[i]\leftarrow m$\;

    \tcp{3) MSA: Predict--Select--Update}
    
    $\hat{\boldsymbol{\pi}}_i \leftarrow \boldsymbol{\pi}_{i-1}\cdot \mathcal{P}$\; \tcp*[r]{Predict (Prior)}

    $(l^\ast,h^\ast)\leftarrow \operatorname*{arg\,max}_{l,h}\;\hat{\boldsymbol{\pi}}_i^{\top}\log\!\big(\mathbf{A}_i^{(l,h)}\big)$\;
    
    $\mathbf{a}^\ast \leftarrow \mathcal{G}_{\sigma}\!\left(\mathbf{A}_i^{(l^\ast,h^\ast)}\right)$\; \tcp*[r]{Select (Observation)}

    $\boldsymbol{\pi}_i \leftarrow \left(\hat{\boldsymbol{\pi}}_i \odot \mathbf{a}^\ast\right)/Z$\; \tcp*[r]{Update (Posterior)}

    \tcp{4) Segment switching via expected aligned position}
    \If{$m < M$ \textbf{\textrm{and}} $\sum_{t=1}^{T} t\cdot \boldsymbol{\pi}_i[t] > b_m$}{
        $m \leftarrow m + 1$ \tcp*[r]{Trigger emotion switch}
    }
}
\Return{$\mathbf{s}$}
\end{algorithm*}

\label{sec:appendix}

\end{document}